\documentclass[11pt]{article}
\pdfoutput=1 
\usepackage{mathtools}
\usepackage{booktabs}
\usepackage[english]{babel}
\usepackage{amsmath,amssymb,amsbsy,amstext, amsthm, simplewick, amsfonts,braket}
\usepackage{graphicx}
\usepackage[small]{caption}
\usepackage{siunitx}
\usepackage{subcaption}
\usepackage{upgreek}
\usepackage{framed}
\usepackage{wrapfig}
\usepackage{multirow}
\usepackage{bbm}
\usepackage[numbers,sort&compress]{natbib}
\usepackage[svgnames,dvipsnames,x11names]{xcolor}
\usepackage[utf8x]{inputenc}
\usepackage{selinput}
\usepackage{bm}
\usepackage{float}
\usepackage{dsfont}
\usepackage[margin=1cm]{caption}
\usepackage{subcaption}
\usepackage{sidecap}
\usepackage{longtable}
\usepackage{anyfontsize}

\usepackage{epstopdf}
\usepackage{cancel}
\usepackage{tcolorbox}
\usepackage{latexsym,amsmath,amssymb,epsfig}
\usepackage{braket}
\usepackage{tensor}
\usepackage{tocloft}
\usepackage[permil]{overpic}

\usepackage{tcolorbox}
\definecolor{greyish2}{rgb}{.96,.96,.96}

\def\xyma{\xymatrix@M.7em}
\def\xymas{\xymatrix@M.1em}

\newcommand{\Comment}[1]{{}}
\definecolor{darkblue}{rgb}{0.15,0.35,0.55}
\definecolor{reddish}{rgb}{0.65, 0.2, 0.2}
\definecolor{darkgreen}{RGB}{50,150,0}
\definecolor{greyish2}{rgb}{.96,.96,.96}
\usepackage[linktocpage=true]{hyperref}
\hypersetup{
colorlinks=true,
citecolor=darkblue,
linkcolor=reddish,
urlcolor=darkblue,
pdfauthor={},
pdftitle={},
pdfsubject={}
}

\DeclareFontFamily{OT1}{rsfs10}{}
\DeclareFontShape{OT1}{rsfs10}{m}{n}{ <-> rsfs10 }{}
\DeclareMathAlphabet{\mathscript}{OT1}{rsfs10}{m}{n}

\newcommand\blfootnote[1]{%
  \begingroup
  \renewcommand\thefootnote{}\footnote{#1}%
  \addtocounter{footnote}{-1}%
  \endgroup
}

\def\gsim{ \lower .75ex \hbox{$\sim$} \llap{\raise .27ex \hbox{$>$}} }
\def\lsim{ \lower .75ex \hbox{$\sim$} \llap{\raise .27ex \hbox{$<$}} }
\def\be{\begin{equation}}
\def\ee{\end{equation}}
\def\bea{\begin{eqnarray}}
\def\eea{\end{eqnarray}}

\def\K{\kappa}
\newcommand{\baaa}{\begin{eqnarray}}
\newcommand{\eaaa}{\end{eqnarray}}

\newcommand{\T}{\mathcal{T}}

\DeclareMathOperator{\E}{e}

\usepackage[a4paper,margin=1in]{geometry}

\usepackage{tikz}
\usetikzlibrary{decorations}
\pgfdeclaredecoration{complete sines}{initial}
{
    \state{initial}[
        width=+0pt,
        next state=upsine,
        persistent precomputation={\pgfmathsetmacro\matchinglength{
            \pgfdecoratedinputsegmentlength / int(\pgfdecoratedinputsegmentlength/\pgfdecorationsegmentlength)}
            \setlength{\pgfdecorationsegmentlength}{\matchinglength pt}
        }] {}
    \state{upsine}[width=\pgfdecorationsegmentlength,next state=downsine]{
        \pgfpathsine{\pgfpoint{0.25\pgfdecorationsegmentlength}{0.5\pgfdecorationsegmentamplitude}}
        \pgfpathcosine{\pgfpoint{0.25\pgfdecorationsegmentlength}{-0.5\pgfdecorationsegmentamplitude}}
    }
    \state{downsine}[width=\pgfdecorationsegmentlength,next state=upsine]{
        \pgfpathsine{\pgfpoint{0.25\pgfdecorationsegmentlength}{-0.5\pgfdecorationsegmentamplitude}}
        \pgfpathcosine{\pgfpoint{0.25\pgfdecorationsegmentlength}{0.5\pgfdecorationsegmentamplitude}}
}
    \state{final}{}
}

\definecolor{greyish}{rgb}{.90,.90,.90}
\definecolor{greyish2}{rgb}{.96,.96,.96}
\usepackage{xcolor,colortbl}
\usepackage{tcolorbox}

\usepackage[all]{xy}

\def \lll {\langle\!\langle}
\def \rr{\rangle\!\rangle}

\newcommand{\p}{\partial}
\renewcommand{\O}{\mathcal{O}}
\renewcommand{\E}{\mathcal{E}}

\renewcommand{\K}{\kappa}
\DeclareSymbolFont{matha}{OML}{txmi}{m}{it}
\DeclareMathSymbol{v}{\mathord}{matha}{118}

\renewcommand{\O}{{\mathcal O}}
\newcommand{\sep}{{\rm sep}}
\newcommand{\bz}{\bar{z}}
\newcommand{\CFT}{{\rm CFT}}
\newcommand{\R}{{\cal R}}

\newcommand{\bE}{\bar{\E}}
\newcommand{\bp}{\bar{\p}}


\numberwithin{equation}{section}
\begin{document}
%
\renewcommand{\thefootnote}{\fnsymbol{footnote}}
\vspace{0truecm}
\thispagestyle{empty}

\begin{center}
{
\bf\LARGE
Null energy constraints on two-dimensional RG flows
}
\end{center}

\vspace{.3truecm}

\begin{center}
{\fontsize{12.7}{18}\selectfont
Thomas Hartman${}^{\rm a}$\blfootnote{\texttt{\href{mailto:hartman@cornell.edu}{hartman@cornell.edu}}}
and Gr\'egoire Mathys${}^{\rm a,b}$\blfootnote{\texttt{\href{mailto:gregoire.mathys@cornell.edu}{gregoire.mathys@cornell.edu}}} 
}

%

\vspace{.8truecm}

\centerline{{\it ${}^{\rm a}$Department of Physics,}}
 \centerline{{\it Cornell University, Ithaca, NY 14850, USA} } 
 
\vspace{.4cm}

\centerline{{\it ${}^{\rm b}$Fields and Strings Laboratory, Institute of Physics}}
 \centerline{{\it Ecole Polytechnique Fédéral de Lausanne (EPFL)} } 
  \centerline{{\it CH-1015 Lausanne, Switzerland} } 

 \vspace{.25cm}

\vspace{.3cm}

\end{center}

\vspace{0.7cm}

\begin{abstract}
\noindent
We study applications of spectral positivity and the averaged null energy condition (ANEC) to renormalization group (RG) flows in two-dimensional quantum field theory. We find a succinct new proof of the Zamolodchikov $c$-theorem, and derive further independent constraints along the flow. In particular, we identify a natural $C$-function that is a completely monotonic function of scale, meaning its derivatives satisfy the alternating inequalities $(-1)^nC^{(n)}(\mu^2) \geq 0$.  The completely monotonic $C$-function is identical to the Zamolodchikov $C$-function at the endpoints, but differs along the RG flow. In addition, we apply Lorentzian techniques that we developed recently to study anomalies and RG flows in four dimensions, and show that the Zamolodchikov $c$-theorem can be restated as a Lorentzian sum rule relating the change in the central charge to the average null energy. This establishes that the ANEC implies the $c$-theorem in two dimensions, and provides a second, simpler example of the Lorentzian sum rule.

\end{abstract}

\newpage

\setcounter{page}{2}
\setcounter{tocdepth}{2}
\tableofcontents
\newpage
\renewcommand*{\thefootnote}{\arabic{footnote}}
\setcounter{footnote}{0}




\section{Introduction}

A fully consistent quantum field theory (QFT) can abstractly be thought of as a renormalization group (RG) flow between two conformal field theories (CFTs). The starting point is a UV fixed point, perturbed by relevant and marginal operators to trigger a flow in the space of theories. It is a longstanding open problem to characterize the allowed flows in theory space, and to understand their underlying structure. 

This subject is guided by $C$-theorems, which are often said to encapsulate the following intuition. Along the renormalization group flow, from the high-energy (UV) fixed point to the low-energy (IR) fixed point, we expect the total number of degrees of freedom to decrease. This is consistent with the Wilsonian picture of renormalization, where high-energy degrees of freedom are progressively integrated out as we proceed to lower energies.  It is however not clear how to characterize degrees of freedom, and $C$-theorems exhibit diverse forms in different dimensions, lacking a unified, dimension-independent description. Moreover, the $C$-theorems in greater than two dimensions have only an indirect and tenuous connection to the intuition that the $C$-function counts degrees of freedom.

A $C$-function is a function on the space of quantum field theories that is universal at conformal fixed points --- that is, it depends only on the physical data of the CFT --- and monotonic under RG flow. If a $C$-function exists, then the renormalization group is irreversible \cite{Zamolodchikov:1986gt}. There is a candidate for a universal quantity that counts degrees of freedom at fixed points, related to the free energy on a sphere  (and to a subleading term in the entanglement entropy) \cite{Cardy:1988cwa,Myers:2010tj,Giombi:2014xxa}. However monotonicity has been proven only in certain cases including dimensions $d=2$ (the $c$-theorem \cite{Zamolodchikov:1986gt}), $d=3$ (the $F$-theorem \cite{Casini:2012ei}), and $d=4$ (the $a$-theorem \cite{Cardy:1988cwa, Komargodski:2011vj}), as well as supersymmetric theories in six dimensions \cite{Heckman:2015axa,Cordova:2015vwa,Cordova:2015fha} and holographic RG flows in all dimensions \cite{Myers:2010xs}. These results highlight a fascinating interplay between holographic duality, quantum information, and more traditional approaches to quantum field theory.

In two dimensions, the degrees of freedom of a CFT are characterized by its central charge, $c$. There are various arguments for why the central charge counts degrees of freedom; for example, it is the coefficient of the thermal free energy \cite{cardyformula} and the vacuum entanglement entropy \cite{Holzhey:1994we}. 
Zamolodchikov's pioneering work identified a particular $C$-function, which is a linear combination of form factors in the stress tensor two point function, and flows to $c$ at conformal fixed points. Invoking reflection positivity of the Euclidean two-point function $\langle \Theta\Theta \rangle$, where $\Theta = T_{\mu}^\mu$ is the trace of the stress tensor, he proved that 
\be 
c_{IR}\leq c_{UV}\, ,
\ee
for the central charges at the endpoints of an RG flow. This is the $c$-theorem.

Our work focuses on deriving universal constraints on two-dimensional QFTs by applying two closely related positivity conditions. The first is the positive spectrum Wightman axiom, which asserts that the spectrum of the momentum operator $P^\mu$ lies in the closed forward lightcone. In Euclidean notation with $ds^2 = |dz|^2$, let us denote the null momenta by
\begin{align}\label{defnull}
\E = P_z , \qquad \qquad \bE = -P_{\bz} \ .
\end{align}
The positive spectrum axiom in two dimensions is equivalent to the statement that these operators are positive semi-definite:
\begin{align}\label{eq:ANECIneq}
\E \geq 0 , \qquad \qquad \bE \geq  0 \ .
\end{align}
The second positivity condition that we will apply is the averaged null energy condition (ANEC). The averaged null energy (ANE)  operator is
\begin{align}
\E_u(v) &=\int du\, T_{uu}(u,v) \ , 
\end{align}
with $u$ a null coordinate. The ANE is a non-local operator that exhibits remarkable properties. In quantum field theory in Minkowski spacetime, the ANEC states that this operator is non-negative,
\begin{align}
\E_u \geq 0 \ .
\end{align}
This was first discussed in the context of general relativity, where it is necessary to prove classic theorems on causality, positive energy, and wormholes \cite{Borde_1987,Gao:2000ga,Graham_2007}. In QFT, it was originally proven in free theories and lower dimensional settings \cite{Klinkhammer:1991ki,Wald:1991xn,Folacci:1992xg,Ford:1995gb}, and there are now two derivations for interacting theories in higher dimensions, that rely respectively on quantum information \cite{Faulkner:2016mzt} and conformal bootstrap techniques \cite{Hartman:2016lgu}. In QFT, the ANEC has provided several non-trivial constraints on the coupling constants and anomaly coefficients in CFTs \cite{Hofman:2008ar,Hofman:2009ug} (see also \cite{Hartman:2016lgu,Hartman:2015lfa,Hartman:2016dxc,Hofman:2016awc,Cordova:2017zej,Cordova:2018ygx,Bautista:2019qxj,Besken:2020snx}). 
It is the prototypical example of a light-ray operator, which have a wide array of applications ranging from holography \cite{Hofman:2008ar,Hofman:2009ug,Kelly:2014mra,Afkhami-Jeddi:2016ntf,Meltzer:2017rtf,Belin:2019mnx,Kologlu:2019bco,Baumann:2019ghk,Belin:2020lsr} to particle phenomenology \cite{Hofman:2008ar,Dixon:2019uzg,Kologlu:2019mfz,Lee:2022ige}, the Lorentzian inversion formula \cite{Hartman:2016lgu,Caron-Huot:2017vep,Simmons-Duffin:2017nub}, and quantum information \cite{Faulkner:2016mzt}.

In two dimensions, the ANEC and the positive spectrum axiom are closely related --- possibly equivalent. As we will discuss below, in a theory that is either conformal or gapped, one can prove $\E_u = P_z$ and so these positivity conditions are identical. It is plausible that this equality holds in a dense set of states in all 2d QFTs, but we do not have a general proof for theories that flow to a nontrivial IR fixed point. We will therefore state our results as derived from either the ANEC or the positive spectrum axiom, keeping in mind that under some additional assumptions these are in fact the same.

Using the positive spectrum axiom, we will first give a very simple new proof of the $c$-theorem. The $C$-function we obtain from this proof is different from Zamolodchikov's, though of course it agrees at the endpoints of the RG flow. While Zamolodchikov's $C$-function is built from a linear combination of $\langle T_{zz}T_{zz}\rangle$, $\langle T_{zz} \Theta\rangle$, and $\langle \Theta\Theta\rangle$, we show that the stress tensor two-point function itself provides a monotonic $C$-function:
\begin{align}
C = 8\pi^2 z^4 \langle T_{zz}(z,\bz)T_{zz}(0)\rangle \ . 
\end{align}
In fact, $C$ is a completely monotonic function of scale, meaning its derivatives along the RG flow have alternating signs. This statement is derived both from the positive spectrum axiom and, separately, from the spectral decomposition of the two-point function, using a theorem relating complete monotonicity to positivity of the inverse Laplace transform. We also find an infinite set of other, similar constraints on the two-point functions.\footnote{We initially found only the first few constraints. We thank Clay C\'ordova for posing the question of whether it is possible to characterize the complete set of constraints of this type.}

We then describe another new derivation of the $c$-theorem, using the Lorentzian method that we developed in a recent paper on the four-dimensional $a$-theorem \cite{Hartman:2023qdn}. This method, based on the ANEC, is suitable for either two or four dimensions as it does not use the connection to the Poincar\'{e} generators, but instead uses the three-point function $\langle \Theta \E_u \Theta\rangle$. In a QFT that flows between two conformal fixed points, we derive the following sum rule for the change in the central charge, written in the metric $ds^2 = -du dv$:
\be 
\Delta c = c_{UV}-c_{IR} = -6\pi\int_{v_1<0}d^2x_1\int_{v_2<0}d^2x_2\, (u_1-u_2)^2\braket{\Theta(x_1)T_{uu}(0)\Theta(x_2)}\, . \label{eq:SumRuleIntro}
\ee
A similar sum rule for the change in the Euler coefficient $\Delta a $ in four dimensions was recently derived in \cite{Hartman:2023qdn} and relies on the same machinery. The only difference between two and four dimensions is the kernel in the integrand. These sum rules come from matching the conformal anomaly in the IR. The sum rule in \eqref{eq:SumRuleIntro} can be rewritten as an expectation value of the averaged null energy,
\begin{align}
\Delta c \approx \langle \psi | \E_u(0) | \psi\rangle \ ,
\end{align}
for a particular wavepacket $|\psi\rangle$ in a low-frequency limit. We therefore obtain the $c$-theorem as a direct consequence of the ANEC in this state. Using the relation between the ANE and the null momentum, we also show that the sum rule \eqref{eq:SumRuleIntro} is equivalent to the sum rule derived by Zamolodchikov. 

\subsubsection*{Outline}

We start with the short and simple derivation of the $c$-theorem in two dimensions from the positivity of null momentum, together with a brief review of the usual derivation, in section \ref{sec:Appetizer}. In section \ref{sec:Monotonicity}, we derive the monotonicity properties that are obeyed by the coefficient functions in $\langle T_{\mu\nu}T_{\alpha\beta}\rangle$. In light of Bernstein's theorem on completely monotonic functions, this implies that some combinations of these form factors have non-negative inverse Laplace transforms, which we confirm using the spectral representation. We then illustrate complete monotonicity in two examples, the free massive scalar and free massive Majorana fermion. In section \ref{sec:AnecIn2d}, we spell out the connection between the ANE operator and the generator of null translations in two dimensions in detail. In section \ref{s:cfromAnec}, we discuss the contact terms at a conformal fixed point in two dimensions, derive the contact terms in the retarded correlator  $\braket{\R\left[\mathcal{E}_u;\Theta\Theta\right]}$ as well as the time-ordered correlator $\braket{\mathcal{T}\left[\mathcal{E}_u\Theta\Theta\right]}$, and match the anomaly between the UV and the IR for a theory that flows between two fixed points. This allows us to derive Lorentzian sum rules that imply the $c$-theorem. We also work out the example of a free massive scalar and free massive Majorana fermion. More details on the computations of the contact term in CFT using the conformal anomaly are presented in appendix \ref{ap:TwoDim}.

\section{A $c$-theorem appetizer\label{sec:Appetizer}}

\subsection{The $c$-theorem from the positive spectrum axiom}

To begin, we will describe what is probably the simplest possible derivation of the Zamolodchikov $c$-theorem. Let us first setup the notation. Our convention for the stress tensor in Euclidean signature is 
\be 
T_{\mu\nu} \equiv \frac{2}{\sqrt{g}}\frac{\delta S_E}{\delta g^{\mu\nu}}\, , \qquad\qquad 
\langle T_{\mu\nu}\rangle = -\frac{2}{\sqrt{g}} \frac{\delta}{\delta g^{\mu\nu}} \log Z\, ,\label{eq:TmunuConv}
\ee
with $S_E$ the Euclidean action. Note that the convention \eqref{eq:TmunuConv} differs by a factor of $-2\pi$ from some of the two-dimensional CFT literature. 
The null momenta defined in \eqref{defnull} are the generators of translations, such that
\be 
[\E, \O(z,\bz)] = i \p \O(z,\bz) , \qquad \qquad  [\bE, \O(z,\bz)] = -i \bar{\p}\O(z,\bz) \, , \label{eq:Poincare}
\ee
with $\partial \equiv \partial_z$ and $\bar{\p} \equiv \partial_{\bar{z}}$.

Let us consider a 2d QFT with mass scale $s =M^2$ that flows between two conformal fixed points. By dimensional analysis, the stress tensor correlator takes the form \cite{Zamolodchikov:1986gt}
\be \label{TTF}
\langle T_{zz}(z,\bz) T_{zz}(0) \rangle = \frac{1}{2(2\pi)^2 z^4} F(z \bz s) \ .
\ee
Inserting the operator $\bar{\mathcal{E}}$, we obtain the following equalities
\be 
s \frac{\p}{\p s} F = \bz \bar{\p} F = -8\pi^2 i \bz z^4 \langle T_{zz}(z,\bz) \bar{\E} T_{zz}(0) \rangle\label{eq:2p4}\, .
\ee 
Choose $z=i$ so that $\langle T_{zz} \bar{\E} T_{zz}\rangle$ can be interpreted as an expectation value, and the prefactor evaluates to $i\bz z^4 = 1$. Applying the positive spectrum axiom \eqref{eq:ANECIneq} we conclude that
\begin{align}\label{dFa}
\frac{\p}{\p s} F \leq 0 \ . 
\end{align}
At the conformal fixed points, $F=c$ and \eqref{dFa} implies the $c$-theorem \cite{Zamolodchikov:1986gt}:
\be 
c_{IR} \leq c_{UV}\, .
\ee
The change in the central charge can also be written as a sum rule. Equation \eqref{eq:2p4} is equivalent to
%
\be 
|z|^2\frac{\partial}{\partial |z|^2}F = -2(2\pi)^2 i |z|^2 z^3\braket{T_{zz}(z,\bar{z}) \bar{\mathcal{E}}T_{zz}(0)}\, .
\ee
Therefore 
\begin{align}
\Delta c = F(0)-F(\infty)= -\int_0^\infty d|z|^2\, \frac{\partial}{\partial |z|^2}F
= 4\pi i \int_{z\neq 0} d^2z\, z^3 \braket{T_{zz}(z,\bar{z}) \bar{\mathcal{E}}T_{zz}(0)}\, .
\end{align}

\subsection{Zamolodchikov's derivation \label{sec:ZamoReview}}
For comparison, let us briefly review Zamolodchikov's original derivation. In addition to \eqref{TTF} the other components of the stress tensor 2-point function take the form 
\begin{align}\label{fgh}
\langle \Theta(z,\bz)T_{zz}(0)\rangle
&= \frac{1}{(2\pi)^2 z^3 \bz }G( s|z|^2) \\
\langle  \Theta(z,\bz) \Theta(0) \rangle
&= \frac{1}{(2\pi)^2z^2 \bz^2} H( s|z|^2) \ , \notag
\end{align}
with similar expressions involving $T_{\bz\bz}$.
%
The trace is $\Theta =T\indices{_\mu^\mu} =  4 T_{z\bz}$, and the conservation equation $\p_\mu T^{\mu\nu} = 0$ implies
\be 
\bar{\p}T_{zz}  = -\frac{1}{4} \p \Theta , \qquad \p T_{\bz\bz} = -\frac{1}{4}\bar{\p}\Theta \ .\label{eq:Conservation}
\ee
This implies the following relations amongst two-point correlators: 
\be 
2\dot{F} + \dot{G} - 3 G = 0\, ,\qquad \qquad  4\dot{G} + \dot{H} - 4 G - 2H = 0\, ,
\ee at separated points ($z\neq 0$), with $\dot{X} =s \frac{\p}{\p s} X $. These relations can be combined into 
\begin{align}\label{zamdot}
s \frac{\p}{\p s}C_{\rm Zam} = -\frac{3}{4}H\, ,
\end{align}
with the Zamolodchikov $C$-function
\begin{align}\label{czam}
C_{\rm Zam} = F-G - \frac{3}{8}H \ .
\end{align}
Reflection positivity requires $H \geq 0$. Therefore, $C_{\rm Zam}(s|z|^2)$ decreases monotonically as a function of $s$.  At long and short distances, the theory is conformal and the trace $\Theta$ vanishes, so $G=H=0$, while $F$ approaches the central charge of the CFT at the fixed points. Therefore 
\be
\lim_{s \to 0} C_{\rm Zam} = F(0) =  c_{UV} , \qquad 
\lim_{s \to \infty} C_{\rm Zam} = F(\infty) = c_{IR} \ . 
\ee
This establishes the $c$-theorem, $c_{UV} \geq c_{IR}$, with strict inequality for nontrivial flows where $\Theta \neq 0$. Integrating \eqref{zamdot} gives the sum rule \cite{Zamolodchikov:1986gt, PhysRevLett.60.2709}
\be 
\Delta c = 3\pi \int_{z \neq 0 } d^2 z\, |z|^2 \langle \Theta(z,\bz) \Theta(0)\rangle \ .\label{eq:ZamoIntegrated}
\ee
Let us compare this original derivation to the derivation above. Positivity of $\langle\Theta\Theta\rangle$ leads to the Zamolodchikov $C$-function $C_{\rm Zam}$, while positivity of $\langle T_{zz}\bE T_{zz}\rangle$ implies that $F$ itself is monotonic, and thus serves as a $C$-function. These are two different $C$-functions that agree at the endpoints, but differ along the flow. The Zamolodchikov $C$-function has a natural flow in the space of couplings, while we will see that $F$ satisfies an infinite set of additional inequalities.

\section{General monotonicity conditions\label{sec:Monotonicity}}

More inequalities can be generated by inserting higher powers of $\E$ and $\bE$ in the stress tensor 2-point functions. We will now show that this leads to an infinite class of constraints on the scale dependence  of the coefficient functions $F$, $G$, and $H$ defined in \eqref{TTF} and \eqref{fgh}.

\subsection{Inequalities from the positive spectrum condition}
 Acting on an arbitrary function $f(|z|^2s)$, we have 
\begin{align}
s\p_s f(|z|^2s) = z\, \p f(|z|^2s) = \bz\, \bp f(|z|^2s) \ . 
\end{align}
Using the Poincare algebra \eqref{eq:Poincare} and applying this identity repeatedly, it is straightforward to derive the relations 
\begin{align}
\langle \Theta(z,\bz) \E^m \bE{}^n \Theta(0) \rangle
&= \frac{1}{(2\pi)^2}\frac{s^{m+2}}{ (iz)^{m+2}(-i\bz)^{n+2} }
\p_s^m \left( s^n \p_s^n \frac{H}{s^2}\right)  \\
\langle T_{zz}(z,\bz) \E^m \bE{}^n T_{zz}(0)\rangle
&= \frac{1}{2(2\pi)^2}\frac{s^{m+4}}{ (-i\bar{z})^n (i z)^{m+4} }\p_s^m \left( s^{n-4} \p_s^n F\right) \ . 
\end{align}
By the positive spectrum axiom, the expectation value $\langle \Psi | \E^m \bE{}^n|\Psi\rangle$ is non-negative in any state $|\Psi\rangle$. By setting $z=i$, $\bz = -i$ we can interpret the two correlation functions above as expectation values in the states $|\Psi\rangle = e^{-\hat{H}/2} \Theta(0)|0\rangle$  and $|\Psi\rangle = e^{- \hat{H}/2} T_{zz}(0)|0\rangle$ respectively, with $\hat{H}$ the Hamiltonian. Accounting for the prefactors we obtain the inequalities
\begin{align}\label{gencon1}
(-1)^{m+n} \p_s^m \left( s^n \p_s^n \frac{H}{s^2}\right) &\geq 0 \\
(-1)^{m+n} \p_s^m \left( s^{n-4} \p_s^n F \right) &\geq 0  \notag\, ,
\end{align}
for any non-negative integers $m$ and $n$. The second inequality,  with $(n,m) = (1,0)$, is the case studied in section \ref{sec:Appetizer} which implies the Zamolodchikov $c$-theorem. Constraints on the coefficient function $G$ can be obtained by using the conservation equation \eqref{eq:Conservation}. For example,
\begin{align}
\E \Theta|0\rangle &= 4 \bE T_{zz}|0\rangle \ , 
\end{align}
which leads to 
\begin{align}
\langle \Theta(z,\bz) \E^{m+1} \bE{}^n\Theta(0)\rangle
&= 
-4 \frac{s^{m+3}}{(2\pi)^2 (iz)^{m+3}(-i\bz)^{n+2}} \p_s^{m} \left( s^{n-1} \p_s^{n+1} \frac{G}{s} \right)  \ .
\end{align}
Setting $z=-\bz = i$ gives the inequalities
\begin{align}\label{gineq}
(-1)^{m+n} \p_s^m \left( s^{n-1} \p_s^{n+1} \frac{G}{s} \right) \geq 0  \ .
\end{align}

\subsection{Complete monotonicity from the spectral measure}

In this section we will show that the generalized inequalities derived above can be restated in terms of completely monotonic functions. The complete monotonicity of $F$ is then analyzed from another perspective in terms of the spectral representation.

\subsubsection{Completely monotonic functions}

A function $f(s)$ is called \textit{completely mononotic} if it is $C^\infty$ on $s \in (0,\infty)$ and its derivatives obey the alternating inequalities
\begin{align}
(-1)^n \frac{d^n}{ds^n} f(s) \geq 0 \ . 
\end{align}
According to the Bernstein theorem \cite{10.1007/BF02592679,de07d37f-965f-344e-9946-a38128703ea8}, a necessary and sufficient condition for complete monotonicity is that the function $f(s)$ is the Laplace transform of a non-negative distribution,
\begin{align}\label{bernstein}
f(s) &= \int_0^\infty d\lambda \, e^{-s\lambda} g(\lambda)\, ,
\end{align}
where $g\geq 0$ and the integral converges for $s>0$. Therefore  according to \eqref{gencon1} and \eqref{gineq} the functions
\begin{align}\label{HnFnGn}
H_n(s,|z|^2) &\equiv (-1)^n s^n \p_s^n \frac{H(s|z|^2)}{s^2}\\
F_n(s,|z|^2) &\equiv  (-1)^n s^{n-4} \p_s^n F(s|z|^2)\notag\\
G_n(s,|z|^2) &\equiv (-1)^n s^{n-1} \p_s^{n+1} \frac{G(s|z|^2)}{s}\notag
\end{align}
are completely monotonic in $s$ for integer $n \geq 0$.

\subsubsection{Complete monotonicity of $F$}

In particular, the case $F_{n=0}$ implies that the function $F(s|z|^2)$, seen as a function of $s$, is completely monotonic. We will now give a second derivation of this fact using the spectral representation to write it as the Laplace transform of a positive function.

The K\"all\'en-Lehmann  spectral representation of the stress tensor 2-point function in two dimensions is \cite{Cappelli:1990yc}
\begin{align}
\langle T_{\mu\nu}(x)T_{\rho\sigma}(0)\rangle
= \frac{1}{12\pi} \int_0^\infty d\mu \,\rho(\mu) \int \frac{d^2 p}{(2\pi)^2}\, e^{ip\cdot x}
\frac{ (g_{\mu\nu}-p_\mu p_\nu)(g_{\rho\sigma}-p_\rho p_\sigma)}{p^2+\mu^2}\, ,\label{eq:StressTensorSpecDensity}
\end{align}
where the spectral density $\rho(\mu)$ is non-negative,
\be 
\rho(\mu) \geq 0 \ . 
\ee
Starting from equation \eqref{eq:StressTensorSpecDensity}, we can perform the Fourier transform to obtain the stress tensor two-point function in terms of a single integral over $\mu$. To perform these integrals, we will need the following standard result:
\be 
\int_0^\infty d\lambda \, \frac{e^{-a^2\lambda -\frac{r^2}{4\lambda}}}{\lambda^{\frac{n}{2}}} = 2^{\frac{n}{2}}\left(\frac{ r}{a}\right)^{1-\frac{n}{2}}K_{\frac{n}{2}-1}\left(a r\right)\, .\label{eq:BesselIntegral}
\ee
We can now start from \eqref{eq:StressTensorSpecDensity} and compute the momentum integral to write the two-point function in terms of a single $\mu$ integral. We derive
\begin{align}
\langle T_{zz}(z)T_{zz}(0)\rangle
&= \frac{2}{12\pi} \int_0^\infty d\mu \,\rho(\mu) \int \frac{dp_z dp_{\bar{z}}}{(2\pi)^2}\, e^{ip_z z + i p_{\bar{z}}\bar{z}}
\frac{ p_z^4}{4p_z p_{\bar{z}}+\mu^2}\label{eq:StressTensorSpecDensity2}\\
 &= \frac{1}{384\pi^2}\left(\frac{\bar{z}}{z}\right)^2\int_0^\infty d\mu\, \rho(\mu) \mu^4 K_4\left(\mu|z|\right)\, ,\label{eq:StressTTrho}
\end{align}
where we introduced a Schwinger parameter $\lambda$ for the denominator, and then used \eqref{eq:BesselIntegral} to perform the $\lambda$ integral.

As shown in \cite{Cappelli:1990yc}, the spectral density takes the general form
\begin{align}
\rho(\mu) = c_{IR} \, \delta(\mu) + \frac{1}{\mu} \hat{\rho}\left( \frac{\mu^2}{s} \right) \ ,\label{eq:RhoFullCF}
\end{align}
where $c_{IR}$ is the central charge of the infrared fixed point, $s$ has dimensions of mass-squared and sets the  scale of the QFT, and $\hat{\rho} \geq 0$. The delta function is defined such that $\int_0^\infty d\mu\, \delta(\mu) = 1$. 
Therefore the form factor is
\begin{align}
F(s|z|^2) &= 8\pi^2 z^4 \langle T_{zz}(z,\bz)T_{zz}(0)\rangle\\
&= c_{IR} + \frac{2\pi}{3}z^4 \int_0^\infty \frac{d\mu}{\mu} \hat{\rho}\left( \frac{\mu^2}{s} \right)
\int \frac{d^2 p}{(2\pi)^2} \frac{ (p_z)^4 e^{ip\cdot x}}{p^2+\mu^2}\\
&=c_{IR}+ \frac{\pi}{3} z^4 \int_0^\infty d\sigma^2\, \sigma^2 \hat{\rho}\left(\sigma^2\right) \int \frac{d^2 p}{(2\pi)^2} \int_0^\infty  d\lambda \, e^{i\sigma  p \cdot x} p_z^4 e^{-\lambda(p^2+s)} \, ,
\end{align}
where $p\cdot x = p_z z + p_{\bz}\bz$. To obtain the third line, we have redefined the integration variables first as $\mu^2\to \sigma^2 s$ and then as $p\to \sigma p$, and introduced a Schwinger parameter for the denominator. 
We can then exchange the order of integration to find the Laplace transform representation
\be 
F(s|z|^2) 
=c_{IR}+  \int_0^\infty d\lambda\, \hat{F}(\lambda, |z|^2) e^{-\lambda s} \label{eq:FfromFhat}, 
\ee
where
\be
\hat{F}(\lambda,|z|^2) 
= \frac{\pi}{3} \int_0^{\infty} \frac{d\sigma^2 }{\sigma^2}
\hat{\rho}\left(\sigma^2\right)
z^4 \p_z^4 \left[
\int \frac{d^2 p}{(2\pi)^2} e^{ip\cdot x \sigma - \lambda p^2}\right]\, .
\ee 
The Gaussian integral is
\be 
\int \frac{d^2p}{(2\pi)^2}e^{i\sigma p\cdot x}e^{-\lambda p^2} = \frac{1}{4\pi\lambda}e^{-\frac{\sigma^2|z|^2}{4\lambda}}\, .
\ee
We can thus finally obtain $\hat{F}(\lambda,|z|^2)$, which is given by 
\begin{align}
\hat{F}(\lambda,|z|^2) 
&= \frac{|z|^8}{3072\lambda^5} \int_0^{\infty} \,d\sigma^2 \,\sigma^6
\hat{\rho}\left(\sigma^2\right)
 e^{-\frac{\sigma^2 |z|^2}{4\lambda}}\label{eq:FhatDef}\, .
\end{align}
This is manifestly positive. It follows that $F(s|z|^2)$ is a completely monotonic function of $s$.

\subsection{Examples}
In this section, we illustrate the generalized monotonicity constraints in the case of a free massive scalar and a free massive Majorana fermion.

\subsubsection{Free massive scalar}
\begin{figure}[t]
    \centering
    \begin{minipage}{0.48\textwidth}
        \centering
        \vspace{0.4cm}
        \includegraphics[width=0.95\textwidth]{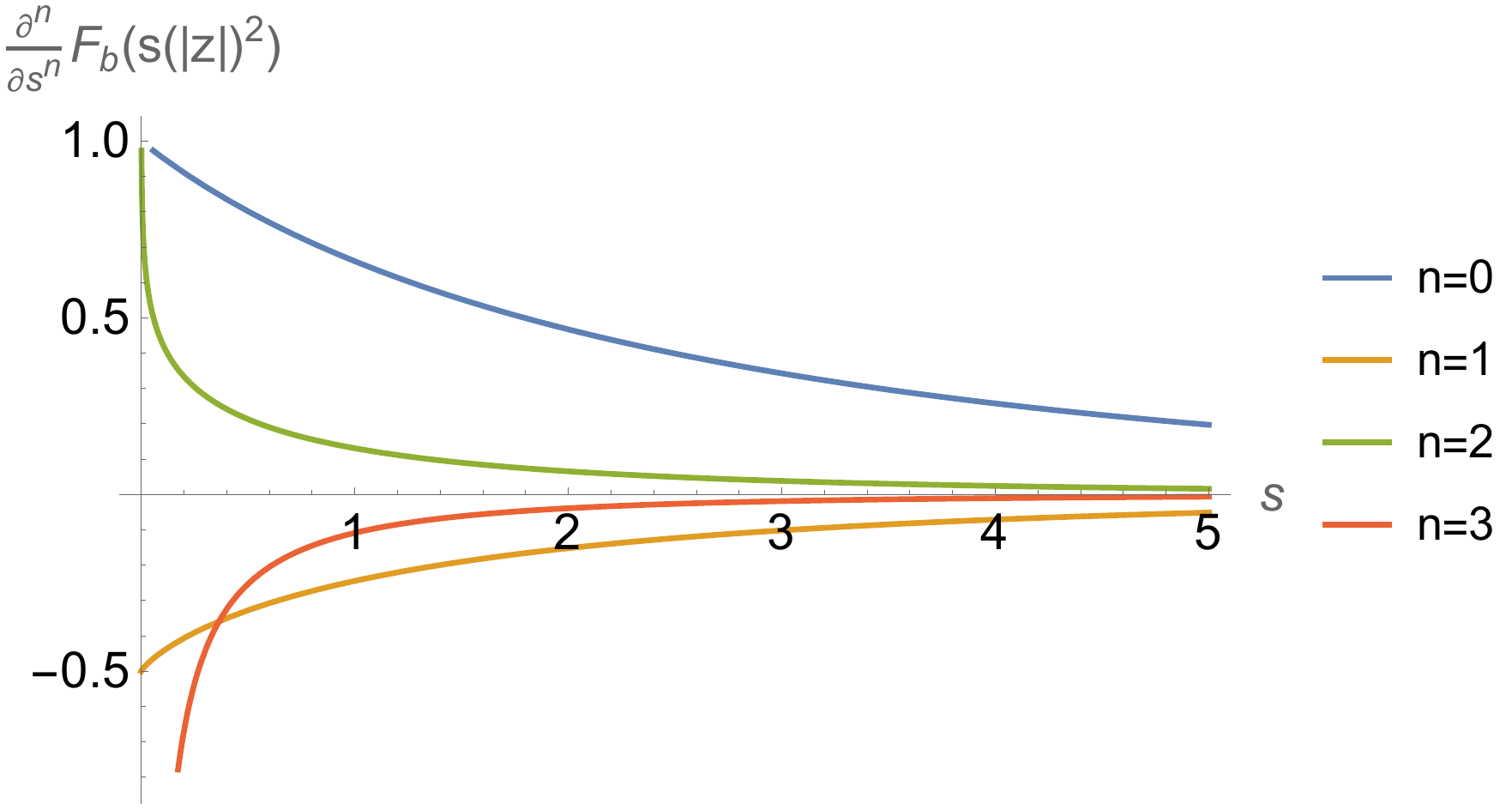} 
        \vspace{0.3cm}
        \caption{The function $F_b\left(s|z|^2\right)$ and its first three derivatives, evaluated at position $z=i,\, \bar{z}=-i$. The derivatives have alternating signs.}
        \label{fig:figureFb}
    \end{minipage}\hfill
    \begin{minipage}{0.48\textwidth}
        \centering
        \includegraphics[width=0.95\textwidth]{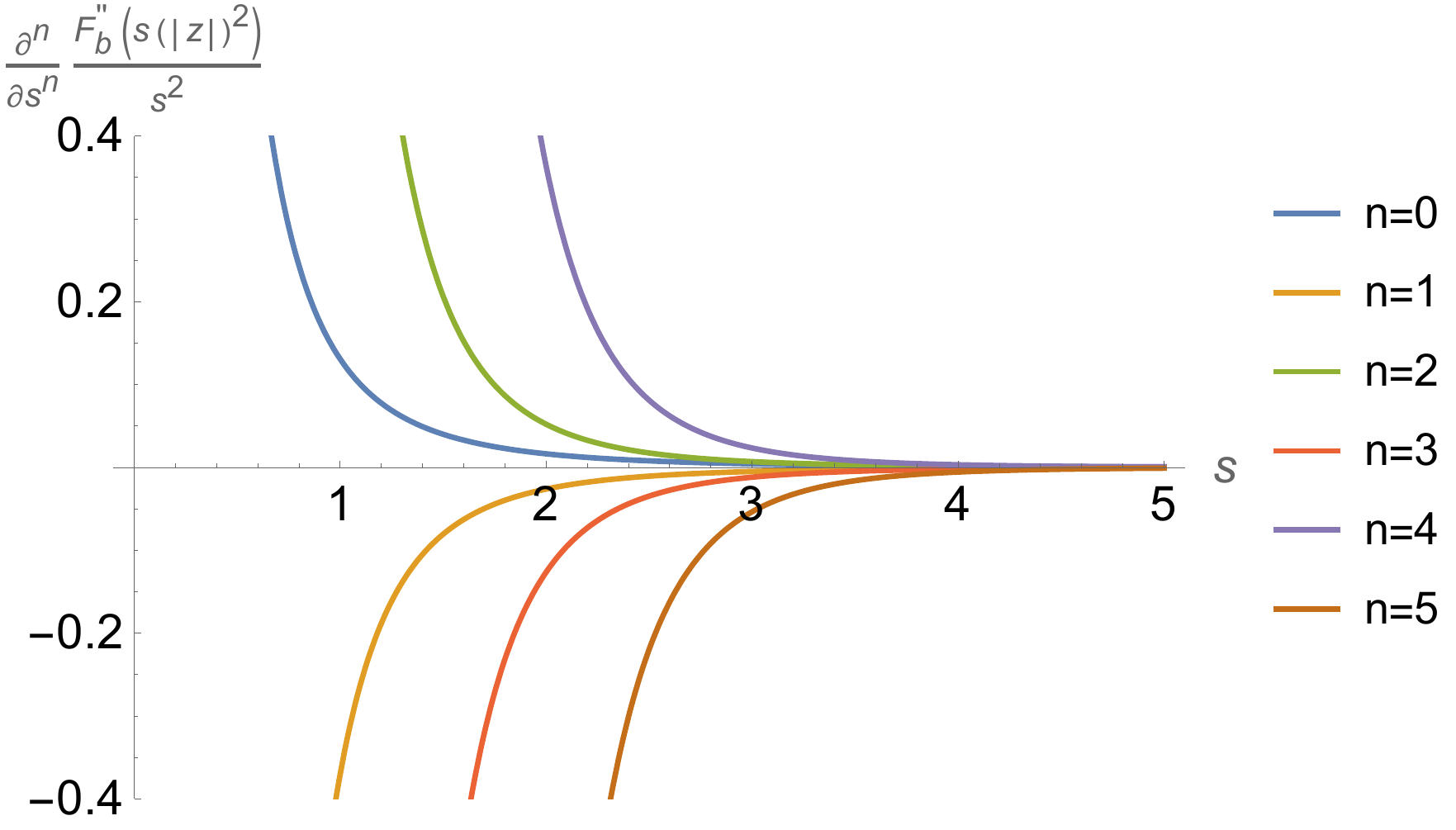} 
        \caption{The function $\frac{F_b''\left(s|z|^2\right)}{s^2}$ and its first five derivatives, evaluated at position $z=i,\, \bar{z}=-i$. The derivatives have alternating signs. }
        \label{fig:FigureFf}
    \end{minipage}
\end{figure}

Consider a free massive scalar field of mass $m$, with the action
\begin{align}
S_b = -\frac{1}{2} \int d^2 x \sqrt{-g}\left( (\p \phi)^2 + m^2 \phi^2\right) \ , \label{eq:bosonAction}
\end{align}
and stress tensor
\begin{align}
T_{\mu\nu} &= -\frac{1}{2}g_{\mu\nu}(m^2\phi^2+(\p \phi)^2)+\p_\mu \phi \p_\nu \phi \ , \label{eq:2dTuv}
\end{align}
such that $T_{zz}(z,\bar{z}) = (\partial_z\phi(z,\bar{z}))^2$. The propagator is
\begin{align}
G_b(z,\bar{z}) = \langle \phi(z,\bar{z})\phi(0) \rangle =  \frac{1}{2\pi} K_0(m|z|)\, ,
\end{align}
where the subscript $b$ refers to boson. The stress tensor two-point function, computed by Wick contractions, is
\begin{align}\label{eq:TTMB}
\langle T_{zz}(z,\bz) T_{zz}(0)\rangle
= 2(-\p_z^2 G_b )^2 
=  \frac{m^4}{32\pi^2}\left(\frac{\bar{z}}{z}\right)^2K_2\left(m\sqrt{z\bar{z}}\right)^2 \ .
\end{align}
The spectral density in the stress tensor two-point function is \cite{Cappelli:1990yc}
\begin{align}
\rho_b(\mu) = 24 \frac{s^2}{\mu^5}\left(1-\frac{4s}{\mu^2}\right)^{-\frac{1}{2}}\theta\left(\mu-2\sqrt{s}\right)\, ,\label{eq:Bosonrho}
\end{align}
with $s=m^2$, which corresponds to
\be 
\hat{\rho}_b(t) = 24\frac{1}{t^2}\left(1-\frac{4}{t}\right)^{-1/2}\theta(t-4)\, .\label{eq:RhoHatMB}
\ee 
Therefore the positive function $\hat{F}$ defined in \eqref{eq:FhatDef} is
\begin{align}
\hat{F}_b(\lambda,|z|^2) 
 &= \frac{|z|^8}{128\lambda^5} \int_4^{\infty} \,d\sigma^2 \,\sigma^2\left(1-\frac{4}{\sigma^2}\right)^{-1/2}
 e^{-\frac{\sigma^2 |z|^2}{4\lambda}}\\
 &= \frac{|z|^6}{16\lambda^5}e^{-\frac{|z|^2}{2\lambda}}\left[|z|^2K_0\left(\frac{|z|^2}{2\lambda}\right)+\left(|z|^2+\lambda\right)K_1\left(\frac{|z|^2}{2\lambda}\right)\right]\, .
\end{align}
Performing the Laplace transform in \eqref{eq:FfromFhat} reproduces the stress tensor form factor obtained from \eqref{eq:TTMB},
\be 
F_b(s|z|^2) = 2(2\pi)^2z^4\braket{T_{zz}(z,\bar{z})T_{zz}(0)} =  \frac{1}{4}s^2 |z|^4 K_2(|z|\sqrt{s})^2 \ ,\label{eq:FMB}
\ee
thus confirming that $F_b$ is a completely monotonic $C$-function. Note that $F_b$ is not the same as the 
Zamolodchikov $C$-function given in \eqref{czam}, though of course it agrees at the endpoints of the flow, where $F_b(0) = c_{UV} = 1$ and $F_b(\infty) = c_{IR} = 0$. The function $F_b$ and its first few derivatives are plotted in figure \ref{fig:figureFb}.  The complete monotonicity of $F_b''/s^2$, which is the case $F_{n=2}$ in \eqref{HnFnGn}, is illustrated in figure \ref{fig:FigureFf}.

\subsubsection{Free massive fermion}

We now consider a free massive Majorana fermion in two dimensions. The action is 
\be 
S_f = \int d^2x \left(\psi \partial_{\bar{z}}\psi  + \bar{\psi}\partial_z\bar{\psi} + im \bar{\psi}\psi\right)\, ,\label{eq:FermionAction}
\ee
where $\psi,\bar{\psi}$ are one-component spinors. 
The propagator is \cite{PhysRevD.15.463}
\begin{align}
\braket{\psi(z,\bar{z})\psi(0,0)} &= 2\partial_z\left[\int \frac{d^2p}{(2\pi)^2}\frac{e^{i p\cdot x}}{p^2 + m^2}\right] =\frac{1}{\pi} \partial_z K_0\left(m |z|\right)  = -\frac{m}{2\pi}\left(\frac{\bar{z}}{z}\right)^{1/2}K_1\left(m|z|\right)\, ,\label{eq:FermionProp1}
\end{align}
and the stress tensor derived from the action \eqref{eq:FermionAction} has
\be 
T_{zz}(z,\bar{z}) =  \frac{1}{2}\psi(z,\bar{z}) \partial_z\psi(z,\bar{z})\, .\label{eq:TzzFermion}
\ee 
Correlators can be computed using Wick's theorem. The stress tensor two-point function is 
\begin{align}
\braket{T_{zz}(z_1,\bar{z}_1)T_{zz}(z_2,\bar{z}_2)} &=\frac{1}{4}\left[\left(\partial_{z_1}\braket{\psi_1\psi_2}\right)\left(\partial_{z_2}\braket{\psi_1\psi_2}\right)-\braket{\psi_1\psi_2}\left(\partial_{z_1}\partial_{z_2}\braket{\psi_1\psi_2}\right)\right]\, ,\label{eq:TzzTzzF2}
\end{align}
where $\psi_i= \psi(z_i,\bar{z}_i)$, leading to
\be 
F_f(s|z|^2) = 2z^4\left[\p_z K_0(\sqrt{s}|z|) \p_z^3 K_0(\sqrt{s}|z|) - (\p_z^2 K_0(\sqrt{s}|z|))^2 \right] \ .\label{eq:FMajorana} 
\ee 
The spectral density for the stress tensor is \cite{Cappelli:1990yc}
\be 
\rho_f(\mu) = 6 \frac{s}{\mu^3}\left(1-\frac{4s}{\mu^2}\right)^{\frac{1}{2}}\theta\left(\mu-2\sqrt{s}\right)\, ,\label{eq:rhof}
\ee
with $s=m^2$, so
\be 
\hat{\rho}_f(t) = 6 \frac{1}{t}\left(1-\frac{4}{t}\right)^{\frac{1}{2}}\theta\left(t-4\right)\, .\label{eq:rhof2}
\ee
Therefore, using \eqref{eq:FhatDef}, $F_f$ is the Laplace transform of the manifestly positive function
\begin{align}
\hat{F}_f(\lambda,|z|^2) 
 &= \frac{|z|^8}{512\lambda^5} \int_4^{\infty} \,d\sigma^2 \,\sigma^4\left(1-\frac{4}{\sigma^2}\right)^{1/2}
 e^{-\frac{\sigma^2 |z|^2}{4\lambda}} \\
 &= \frac{|z|^4}{32\lambda^4}e^{-\frac{|z|^2}{2\lambda}}\left[|z|^2K_0\left(\frac{|z|^2}{2\lambda}\right)+\left(|z|^2+4\lambda\right)K_1\left(\frac{|z|^2}{2\lambda}\right)\right]\, . \notag
\end{align}
It follows that $F_f$ is a completely monotonic $C$-function, interpolating from $F_f(0) = c_{UV} = \frac{1}{2}$ in the UV to $F_f(\infty) = c_{IR} = 0$ in the IR.

\section{The ANEC in two dimensions\label{sec:AnecIn2d}}

So far we have worked in Euclidean signature and derived constraints from the positive spectrum axiom. We will now venture into Lorentzian signature, with the aim of writing a Lorentzian sum rule for $\Delta c$ and deriving the $c$-theorem from the averaged null energy condition (ANEC). We will first discuss the close relation between the ANEC and the positive spectrum axiom.

In two-dimensional Minkowski space we use null coordinates $ds^2 = -du dv$, with $u=t-y$, $v=t+y$. These are related by analytic continuation to the Euclidean coordinates $ds^2 = |dz|^2$ by $z=-u, \bz = v$. 
With our conventions, the stress tensor is $T_{\mu\nu} = -\frac{2}{\sqrt{-g}} \frac{\delta S}{\delta g^{\mu\nu}}$, and the classical null energy condition is $T_{uu}, T_{vv} \geq 0$. The momentum operator is
\begin{align}
P_\mu &= - \int_{\Sigma} \sqrt{h} T_{\mu\nu}u^\nu\, ,
\end{align}
where $\Sigma$ is a Cauchy slice and $u^\mu$ is the forward-pointing timelike unit normal.

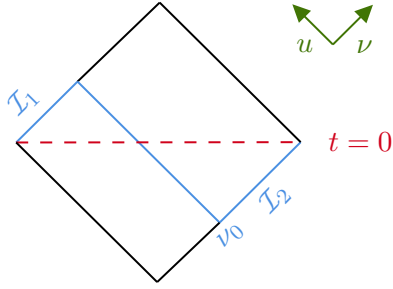
\begin{figure}[t]
\begin{center}
\tikzset{every picture/.style={line width=0.75pt}} 
\begin{tikzpicture}[x=0.75pt,y=0.75pt,yscale=-1,xscale=1]
\draw [color={rgb, 255:red, 74; green, 144; blue, 226 }  ,draw opacity=1 ][fill={rgb, 255:red, 7; green, 52; blue, 186 }  ,fill opacity=1 ]   (89.65,85.03) -- (120.28,54.4) ;
\draw    (120.28,54.4) -- (161.29,14.65) ;
\draw [color={rgb, 255:red, 74; green, 144; blue, 226 }  ,draw opacity=1 ]   (120.28,54.4) -- (191.21,125.33) ;
\draw [color={rgb, 255:red, 74; green, 144; blue, 226 }  ,draw opacity=1 ]   (191.21,125.33) -- (231.67,84.87) ;
\draw [color={rgb, 255:red, 208; green, 2; blue, 27 }  ,draw opacity=1 ] [dash pattern={on 4.5pt off 4.5pt}]  (89.65,85.03) -- (231.67,84.87) ;
\draw    (161.29,14.65) -- (231.67,84.87) ;
\draw    (89.65,85.03) -- (160.03,155.25) ;
\draw    (160.03,155.25) -- (191.21,125.33) ;
\draw [color={rgb, 255:red, 65; green, 117; blue, 5 }  ,draw opacity=1 ]   (247.75,35.75) -- (229.62,17.62) ;
\draw [shift={(227.5,15.5)}, rotate = 45] [fill={rgb, 255:red, 65; green, 117; blue, 5 }  ,fill opacity=1 ][line width=0.08]  [draw opacity=0] (8.93,-4.29) -- (0,0) -- (8.93,4.29) -- cycle    ;
\draw [color={rgb, 255:red, 65; green, 117; blue, 5 }  ,draw opacity=1 ]   (247.75,35.75) -- (265.63,17.87) ;
\draw [shift={(267.75,15.75)}, rotate = 135] [fill={rgb, 255:red, 65; green, 117; blue, 5 }  ,fill opacity=1 ][line width=0.08]  [draw opacity=0] (8.93,-4.29) -- (0,0) -- (8.93,4.29) -- cycle    ;
\draw (244,78.4) node [anchor=north west][inner sep=0.75pt]    {$\textcolor[rgb]{0.82,0.01,0.11}{t=0}$};
\draw (81.79,64.76) node [anchor=north west][inner sep=0.75pt]  [rotate=-315]  {$\textcolor[rgb]{0.29,0.56,0.89}{\mathcal{I}_{1}}$};
\draw (207.33,113.99) node [anchor=north west][inner sep=0.75pt]  [rotate=-314.07]  {$\textcolor[rgb]{0.29,0.56,0.89}{\mathcal{I}_{
2}}$};
\draw (186.76,133.3) node [anchor=north west][inner sep=0.75pt]  [color={rgb, 255:red, 74; green, 144; blue, 226 }  ,opacity=1 ,rotate=-317.38]  {$\nu _{0}$};
\draw (228,31.9) node [anchor=north west][inner sep=0.75pt]    {$\textcolor[rgb]{0.25,0.46,0.02}{u}$};
\draw (258,32.4) node [anchor=north west][inner sep=0.75pt]    {$\textcolor[rgb]{0.25,0.46,0.02}{\nu }$};
\end{tikzpicture}
\caption{Relation between the null translation generator $P_u$ and the ANE operator $\E_u$ in 2d QFT. The null momentum $P_u$, initially defined on the $t=0$ slice, is deformed to the blue contour. This implies $P_u = -\E_u$ so long as $\Theta$ falls off fast enough that the contributions along ${\cal I}_1$ and ${\cal I}_2$ vanish.}\label{fig:EvsP}
\end{center}
\end{figure}

The averaged null energy (ANE) is the operator
\begin{align}
\E_u(v) = \int du\, T_{uu}(u,v) \ .
\end{align}
This operator has several nice properties, but in particular, the ANE operator is non-negative:
\be \langle \psi |\E_u(v)|\psi\rangle \geq 0\, ,\label{eq:ANEC}
\ee
in any state \cite{Klinkhammer:1991ki,Wald:1991xn,Folacci:1992xg,Ford:1995gb,Faulkner:2016mzt,Hartman:2016lgu}.  In two dimensions, the ANEC is almost equivalent to the positive spectrum axiom, which states that the spectrum of the momentum operator lies in the closed forward cone \cite{Haag:1992hx}:
\begin{align}
P^t \geq 0 , \qquad P^2 \leq 0 \ .
\end{align}
Equivalently, 
\begin{align}
P_u \leq 0 , \qquad P_v \leq 0 \ .
\end{align}
To relate this to the ANEC, let us write the null momentum as an integral over the $t=0$ Cauchy slice,
\begin{align}
P_u &= - \int_{-\infty}^\infty dy T_{ut} \ . 
\end{align}
Deforming the contour of integration as shown in figure \ref{fig:EvsP} gives 
\begin{align}\label{pWithTrace}
P_u &= -\E_u(v_0) + \frac{1}{4}\int_{{\cal I}_1 \,\cup \,{\cal I}_2} dv\, \Theta \ , 
\end{align}
where we used $T_{vu} = -\tfrac{1}{4}\Theta$. ${\cal I}_1$ and ${\cal I}_2$ are segments of future and past null infinity, respectively:
\begin{align}
{\cal I}_1 = \{u=\infty, v\in (-\infty,v_0]\} \, \qquad
{\cal I}_2 = \{u=-\infty, v\in [v_0,\infty)\} \ .
\end{align}
In states satisfying the physical condition that there is no flux of right-moving energy along ${\cal I}_1$ or  ${\cal I}_2$, the trace term in \eqref{pWithTrace} vanishes and we have
\begin{align}\label{EPrel}
\E_u = -P_u \geq 0  \ . 
\end{align}
It seems plausible that $\E_u = -P_u$ holds within a dense set of states in any QFT, but we do not have a general proof. In a CFT, it holds because $\Theta = 0$. In a theory with a mass gap, a rigorous proof of \eqref{EPrel} in algebraic QFT can be found in \cite[Theorem 2.5]{Verch:1999nt}. For theories with a flow to a nontrivial IR fixed point, with local operators inserted at finite distance from the origin, it also holds for the leading terms in the OPE. 

If we assume $\E_u = -P_u$, then the ANEC follows from the positive spectrum axiom. This is quite different from the situation in higher dimensions, where the derivation of the ANEC in interacting QFT relies on monotonicity of relative entropy \cite{Faulkner:2016mzt} or causality/analyticity of correlation functions \cite{Hartman:2016lgu}. 

\section{Lorentzian sum rule\label{s:cfromAnec}}

Consider a 2d QFT with mass scale $M$ that flows from CFT$_{\rm UV}$ to CFT$_{\rm IR}$. Connected correlation functions in the QFT are denoted $\langle \cdot\rangle$, while the connected correlation functions of the UV and IR CFTs are denoted $\langle \cdot\rangle_{\rm UV}$ and $\langle \cdot\rangle_{\rm IR}$. When not referring to any particular CFT, the CFT correlators are written $\langle \cdot \rangle_{\rm CFT}$.

The trace of the stress tensor in a 2d CFT in curved space is
\begin{align}\label{traceCFT}
\langle \Theta \rangle_{\rm CFT} = \frac{c}{24\pi}R + \frac{b}{24\pi}\Lambda^2 \ .
\end{align}
The first term is the Weyl anomaly, with $c$ the central charge and $R$ the Ricci scalar of the background spacetime. The second term, with $\Lambda$ the UV cutoff and $b$ a dimensionless constant, can be removed by a local counterterm so it is customarily set to zero. It corresponds to a cosmological constant in the effective action.

We tune the cosmological constant to zero in the UV, setting $b_{UV} = 0$. The cosmological constant will be generated along the RG flow, so in order to match the QFT we write the trace at the UV and IR fixed points as
\begin{align}
\langle \Theta\rangle_{\rm UV} &= \frac{c_{UV}}{24\pi}R \\
\langle \Theta\rangle_{\rm IR} &= \frac{c_{IR}}{24\pi}R + \frac{b}{24\pi}M^2 \ .
\end{align}
In this section we will discuss two manifestly-positive sum rules for the difference in central charge between the two CFTs, i.e. $\Delta c = c_{UV} - c_{IR}$. The first is Zamolodchikov's sum rule relating $\Delta c$ to the Euclidean 2-point function, $\langle \Theta \Theta\rangle$, which we review. We then derive the new sum rule relating $\Delta c$ to the averaged null energy $\langle \Theta \E_u \Theta\rangle$ using the method developed in \cite{Hartman:2023qdn}. We also show that in two dimensions, these sum rules are in fact equivalent using the identity $\E_u = -P_u$ to reduce the 3-point function to a 2-point function.

\subsection{Review of the Zamolodchikov sum rule}
In section \ref{sec:ZamoReview} we reviewed Zamolodchikov's derivation of the $c$-theorem from the conservation laws, which only uses the correlation functions at separated points. We will now review how the same sum rule can be obtained by the studying the contact term in $\langle\Theta\Theta\rangle$.

In a CFT, any stress tensor correlation function involving at least one $\Theta$ can be obtained by varying the Weyl anomaly \eqref{traceCFT} with respect to the metric. This is reviewed in appendix \ref{sec:CorrelatorConventions}. The first variation gives the Euclidean 2-point function, which is 
\begin{align}
\langle \Theta(x_1)\Theta(x_2)\rangle_{\rm CFT}
&= \frac{1}{12\pi}(-c\p^2 + b \Lambda^2)\delta^{(2)}(x_1-x_2) \, , \label{eq:EuclidenTraceTrace}
\end{align}
and is derived in detail in appendix \ref{ap:TwoDim}.
The Fourier transform is
\begin{align}
\lll \Theta(K)\Theta(-K)\rr_{\rm CFT} = \int d^2 x \, e^{iK\cdot x}\langle \Theta(x)\Theta(0)\rangle_{\rm CFT}
&= \frac{1}{12\pi}( cK^2 + b \Lambda^2) \ ,\label{eq:TINv}
\end{align}
where we use the double bracket notation, 
\be 
\braket{\mathcal{O}(k_1)\cdots\mathcal{O}(k_n)} \equiv (2\pi)^d \delta^{(d)}(k_1+\cdots+k_n)\lll \mathcal{O}(k_1)\cdots\mathcal{O}(k_n)\rr\, .\label{eq:DBracket}
\ee
This computation is presented in appendix \ref{ap:TwoDim}.
The formula \eqref{eq:TINv} can be inverted to solve for the central charge $c$. We simply act on both sides with the momentum Laplacian $\p_K^2 = \p_{K^1}^2 + \p_{K^2}^2$, then set the momentum to zero. This produces the identity
\begin{align}\label{cCFT}
c = 3\pi \left.\p_{K}^2\right|_{K=0} \lll \Theta(K)\Theta(-K)\rr_{\rm CFT}  = -3\pi \int d^2 x \, x^2 \langle \Theta(x)\Theta(0)\rangle_{\rm CFT} \ .
\end{align}
The integral in \eqref{cCFT} is over Euclidean space, and the entire contribution comes from the contact term at $x=0$. 

We will now apply this formula to the infrared CFT coming from an RG flow. The correlators of the QFT match those of  CFT$_{IR}$ at low momentum, i.e.
\begin{align}
\langle \Theta(K) \Theta(-K)\rangle \approx \langle \Theta(K)\Theta(-K)\rangle_{\rm IR}\, ,
\end{align}
for $K^2 \ll M^2$. The match includes the $O(K^2)$ term that is responsible for the IR anomaly. Therefore,
\begin{align}
c_{IR} 
&= 3\pi \left.\p_{K}^2\right|_{K=0} \lll \Theta(K)\Theta(-K)\rr\, , 
\end{align}
where now the correlator on the right-hand side is in the QFT, not the CFT. Writing this as a Fourier integral, we obtain
\begin{align}\label{cmid}
c_{IR} &= -3\pi \int d^2 x \, x^2 \langle \Theta(x)\Theta(0)\rangle \ . 
\end{align}
The integral has a UV contact term at $x=0$, which is controlled by the UV CFT, plus contributions from separated points. Let us write the 2-point function in the QFT as
\begin{align}
\langle \Theta(x)\Theta(0)\rangle = \langle \Theta(x)\Theta(0)\rangle_{\rm UV} + \langle \Theta(x)\Theta(0)\rangle_\sep \ , 
\end{align}
where the first term is a pure contact term, and the second term is only nonzero at separated points. Moving the UV contact term to the other side of \eqref{cmid}, we find the sum rule
\begin{align}\label{zamosum}
\Delta c \equiv c_{UV} - c_{IR} = 3\pi \int d^2 x \, x^2 \langle \Theta(x)\Theta(0)\rangle_\sep \ .
\end{align}
This is the sum rule of Zamolodchikov \cite{Zamolodchikov:1986gt} (see also \cite{PhysRevLett.60.2709,Cappelli:1990yc,Komargodski:2011xv}), as reviewed in section \ref{sec:ZamoReview}. The right-hand side is manifestly positive because in Euclidean signature, reflection positivity implies that 
\begin{align}
\langle \Theta(x)\Theta(0)\rangle_\sep \geq 0 \, .
\end{align}

\subsection{Sum rule from the 3-point function}
We will now apply a similar strategy to the correlation function $\langle\Theta \E_u\Theta\rangle$, following \cite{Hartman:2023qdn}. This is a Lorentzian correlation function, so ordering is important, and the sum rule will be in Lorentzian signature. We nonetheless start in Euclidean signature. Varying the Weyl anomaly a second time gives the Euclidean 3-point function $\langle \Theta\Theta T_{\alpha\beta}\rangle_{\rm CFT}$. In momentum space, the result for the null-null component (which is presented in appendix \ref{ap:TwoDim}) is
\begin{align}\label{theta3e}
\lll \Theta(K_1)\Theta(K_2) T_{uu}(K_3) \rr_{\rm CFT} = \frac{c}{6\pi} K_{1u}K_{2u} \ .
\end{align}
Since this is a pure contact term, the analytic continuation to the time-ordered, anti-time-ordered, retarded, or advanced correlator in Lorentzian signature is trivial in momentum space, as reviewed in \cite{Meltzer:2021bmb, Hartman:2023qdn}. Time ordering $\T$ is defined in the standard way, and the retarded ordering is defined
\begin{align}
\langle \R[ T_{\alpha\beta}(x_3); \O(x_1)\O(x_2)]\rangle
&= -\theta(t_3-t_1)\theta(t_1-t_2) \langle [[T_{\alpha\beta}(x_3), \O(x_1)], \O(x_2)]\rangle - (x_1 \leftrightarrow x_2) \notag\\
&\qquad  + \mbox{contact terms} \ . 
\end{align}
For $\O = \Theta$ in a CFT only the contact terms are nonzero. 
By analytically continuing \eqref{theta3e} we obtain
\begin{align}\label{R32d}
\lll \R[T_{uu}(k_3); \Theta(k_1)\Theta(k_2)]\rr_\CFT
= - \lll {\cal T}[\Theta(k_1)\Theta(k_2)T_{uu}(k_3)]\rr_\CFT 
= \frac{ c}{6\pi} k_{1u}k_{2u} \ . 
\end{align}
We use $K$ for Euclidean momentum and $k$ for Lorentzian momentum.
The ANE operator is
\begin{align}\label{Eukuv}
\E_u(v) = \frac{1}{\pi} \int dk_v \,e^{ik_v v} T_{uu}(k_u = 0, k_v) = \int du\, T_{uu}(u,v) \ .
\end{align}
This relation is used to calculate retarded and time-ordered correlators of the ANE from $\langle \Theta\Theta T_{\alpha\beta}\rangle$. Thus integrating \eqref{R32d} using $\delta^{(2)}(k) = \frac{1}{2}\delta(k_u) \delta(k_v)$ we find
\begin{align}\label{anecR2}
\langle \R[\E_u(0); \Theta(k_1)\Theta(k_2)]\rangle_\CFT = -\langle \mathcal{T}[\E_u(0) \Theta(k_1)\Theta(k_2)]\rangle_\CFT
&= \frac{c}{3}  k_{1u} k_{2u}\delta(k_{1u}+k_{2u}) \ .
\end{align}
Note that $\E_u(0) = \E_u(v=0)$ is in position space while the other operators are in momentum space.
The equation \eqref{anecR2} can be inverted to solve for $c$ as follows:
Write the retarded correlator in \eqref{anecR2} as a Fourier transform, integrate both sides $\int dk_{2u}$, act with $-\frac{3}{2}\p_{k_{1u}}^2$, then set the momenta to zero. 
The result is 
\begin{align}\label{c2fixu2}
c = 3\pi \int d^2 x_1 \int d^2x_2 \,  u_1^2  \delta(u_2) \langle \R[\E_u(0); \Theta(x_1)\Theta(x_2)]\rangle_\CFT \ .
\end{align}
This holds in CFT, where the correlator is a pure contact term. To turn it into an RG sum rule we will now follow the same strategy as in the derivation of the Zamolodchikov sum rule above. First, we apply \eqref{c2fixu2} to the IR CFT. Since the integral is a correlation function at zero momentum, we can replace the IR CFT correlation function by the QFT correlation function,
\begin{align}\label{cir1}
c_{IR} = 3\pi \int d^2 x_1 \int d^2x_2 \,  u_1^2  \delta(u_2) \langle \R[\E_u(0); \Theta(x_1)\Theta(x_2)]\rangle  \ .
\end{align}
Now we
split the QFT correlator into three pieces:
\begin{align}\label{Rsplit}
\langle \R[\E_u(0); \Theta(x_1)\Theta(x_2)]\rangle
&=
\langle \R[\E_u(0); \Theta(x_1)\Theta(x_2)]\rangle_\sep
+
\langle \R[\E_u(0); \Theta(x_1)\Theta(x_2)]\rangle_{PC}\\
&\qquad 
+
\langle \R[\E_u(0); \Theta(x_1)\Theta(x_2)]\rangle_{UV} \notag\, ,
\end{align}
which are the terms with no points coincident (labeled `sep' for separated), two points coincident (labeled `PC' for partial contact), and three points coincident, respectively.\footnote{The formula \eqref{Rsplit} must be interpreted carefully: It holds when integrated against a test function such that all three terms converge individually. Otherwise, the split into three terms can have ambiguities. For example, if the separated term is highly singular, then its Fourier transform diverges and must be regulated, and the choice of regulator affects the contact terms.}
In \cite{Hartman:2023qdn} we showed that in $d \leq 4$, partial contact terms in this correlation function can only come from marginal spin-2 operators $\O_{\alpha\beta}$ other than the stress tensor. In two dimensions, assuming an interacting theory so there is a unique stress tensor, there are no such operators so the partial contact term vanishes. This is a special feature of the lightray correlation function, as the derivation of this result in \cite{Hartman:2023qdn} uses both the fact the the ANE operator annihilates the vacuum $\E_u|0\rangle = 0$ \cite{Epstein:1965zza} and properties of the stress tensor OPE.

Applying \eqref{Rsplit} to \eqref{cir1}, discarding the vanishing partial contact term, and moving the UV contact term to the other side, we obtain the sum rule
\begin{align}
c_{UV} - c_{IR} &= -3\pi \int d^2 x_1 \int d^2x_2 \,  u_1^2  \delta(u_2) \langle \R[\E_u(0); \Theta(x_1)\Theta(x_2)]\rangle_{\rm sep} \ . 
\end{align}
The retarded correlator at separated points is by definition
\begin{align}
 \langle \R[\E_u(0); \Theta(x_1)\Theta(x_2)]\rangle_{\rm sep}
 &= -\theta(-v_1)\theta(v_1-v_2)\langle [[ \E_u(0), \Theta(x_1)],\Theta(x_2)]\rangle_\sep
   - (1\leftrightarrow 2) \ . 
\end{align}
When the nested commutator is expanded in Wightman functions, many of the terms vanish using $\E_u|0\rangle = 0$. Only the two orderings where each trace is on one side of the lightray integral (i.e. $\langle \Theta \E_u \Theta\rangle$) survive. We therefore obtain the sum rule
\begin{align}\label{csumR}
c_{UV} - c_{IR} &= -6\pi \int_{v_1<0}d^2 x_1 \int_{v_2<0}d^2 x_2 \, 
u_1^2 \delta(u_2)\langle \Theta(x_1)\E_u(0) \Theta(x_2)\rangle\, .
\end{align}
There is no contribution to the integral from coincident points. 

In \eqref{csumR} the $u_2$ position is fixed to zero. Using translation invariance we can instead fix the null energy to the origin, which leads to the equivalent, more symmetrical sum rule
\begin{align}\label{csumTuu}
c_{UV} - c_{IR} &= -6\pi \int_{v_1<0}d^2 x_1 \int_{v_2<0}d^2 x_2 \, 
(u_1-u_2)^2 \langle \Theta(x_1)T_{uu}(0) \Theta(x_2)\rangle \ . 
\end{align}

If we had started from the time-ordered rather than retarded correlator, the only differences would be the sign of the contact term and the arguments of the step functions when the ordered correlator is expanded in Wightman functions. This leads to a time-ordered sum rule
\begin{align}\label{csumT}
c_{UV} - c_{IR} &= 6\pi \int_{v_1>0}d^2 x_1 \int_{v_2<0}d^2 x_2 \, 
u_1^2 \delta(u_2)\langle \Theta(x_1)\E_u(0) \Theta(x_2)\rangle \ ,
\end{align}
which is equivalent to \eqref{csumR} under rotating the $v_1$ contour.

\subsubsection{Comparison to the Zamolodchikov sum rule}

We will now show this is equivalent to the Zamolodchikov sum rule \eqref{zamosum}. 
Using 
\begin{align}
[\E_u, \O] = -[P_u, \O] = -i \p_u \O\, ,
\end{align}
and $d^2x = \frac{1}{2}du dv$, the sum
 rule \eqref{csumT} becomes
\begin{align}
c_{UV} - c_{IR} &= \frac{3\pi}{2} i  \int du_1 dv_1 dv_2\,  \theta(v_1)\theta(-v_2) u_1^2 \p_{u_1}\langle \Theta(u_1,v_1)\Theta(0,v_2)\rangle_{\rm sep} \\
&=
- 3\pi i \int du_1 dv_1 dv_2 \,\theta(v_1)\theta(-v_2)  u_1 \langle \Theta(u_1, v_1-v_2)\Theta(0)\rangle_{\rm sep}\\
&= 6\pi i \int d^2 x\, \theta(v)x^2 \langle \Theta(x)\Theta(0)\rangle_{\rm sep} \\
&= 3\pi i \int d^2 x\, x^2 \langle {\cal T}[\Theta(x)\Theta(0)]\rangle_{\rm sep} \, .
\end{align}
We integrated by parts in the second line and used $\int_0^{\infty} dv_1 \int_{-\infty}^{0} dv_2\, G(v_1-v_2)
 = \int_{0}^{\infty} dv\, v \,G(v)$ and $x^2 = -uv$ in the third line. These formulas are all in Minkowski space. Wick rotating the last line gives the Zamoldchikov formula \eqref{zamosum}.

\subsection{The $c$-theorem from the ANEC}
The sum rules in \eqref{csumR}-\eqref{csumT} are not manifestly positive. To prove the $c$-theorem from the ANEC, we must write $\Delta c=c_{UV} - c_{IR}$ in terms of the expectation value of $\E_u$ in some state. 
The obvious candidate from \eqref{csumR} is a state created by an insertion of the trace in the region $v<0$. Consider the wavepacket
\begin{align}
|\Psi(k_u)\rangle &= \int d^2 x\, \theta(-v) e^{ik_u u - \frac{u^2}{\sigma^2}}\Theta(u,v)|0\rangle
\end{align}
where $\sigma$ is an infrared cutoff with $\sigma^{-1}\ll k_u \ll M$. The sum rule \eqref{csumR} is equivalent to the statement that at leading order in $k$, 
\begin{align}
\int_{v_1<0}d^2x_1 \int_{v_2<0}d^2 x_2  e^{ik_1\cdot x_1+ik_2 \cdot x_2}\langle\Theta(x_1)\E_u(0)\Theta(x_2)\rangle
&\approx \frac{c_{UV} - c_{IR}}{6} k_{1u}^2 \delta(k_{1u}+k_{2u}) \ .
\end{align}
The expectation value of the ANE in the state $|\Psi(k_u)\rangle$ is calculated from this relation by a convolution with the Gaussian damping factor. The result at order $k^2$ is
\begin{align}
\langle \Psi(k_u)| \E_u(0) | \Psi(k_u)]\rangle \approx (c_{UV} - c_{IR}) k_u^2 \frac{\sigma}{12\sqrt{2\pi}} \ ,
\end{align}
with corrections suppressed by $k_u/M$ and $1/(k_u \sigma)$. Thus the ANEC, $\langle \Psi(k_u)| \E_u(0) | \Psi(k_u)]\rangle \geq 0$, implies the $c$-theorem $c_{UV} \geq c_{IR}$.

\subsection{Examples}

In this section, we will illustrate the ANE sum rule for a free massive scalar field and a free massive fermion.

\subsubsection{Free massive scalar}
We will first apply the ANE sum rule to a free massive scalar field in two dimensions, which flows from the massless scalar in the UV with $c_{UV} = 1$ to the trivial theory in the IR with $c_{IR} = 0$. The analogous calculation in four dimensions was described in \cite{Hartman:2023qdn}. 

The action for a free massive scalar is given in \eqref{eq:bosonAction}, while the stress-tensor is presented in \eqref{eq:2dTuv}. From this stress-tensor, we can obtain the trace and null energy, which are
\begin{align}
\Theta = -m^2 \phi^2 \ , \qquad\qquad T_{uu} = (\p_u\phi)^2 \ . 
\end{align}
The Feynman propagator  is
\begin{align}
G_b(x-y) = \langle \T[\phi(x)\phi(y)]\rangle &= -i \int \frac{d^2 p}{(2\pi)^2} \frac{e^{ip\cdot(x-y)}}{p^2+m^2-i\epsilon} \ . 
\end{align}
The 3-point function at separated points, calculated by Wick contractions, is
\begin{align}
\langle \T[ T_{uu}(x_3)\Theta(x_1)\Theta(x_2)]\rangle_\sep = 8m^4 G_b(x_1-x_2) \p_{u_1}\p_{u_2}\left[ G_b(x_1-x_3)G_b(x_2-x_3)\right] \ . 
\end{align}
In momentum space, 
\begin{align}\label{tttsep2dLoop}
&\lll {\cal T}[ T_{uu}(k_3)\Theta(k_1)\Theta(k_2)]\rr_{\rm sep} = \\
&\qquad 8m^4i \int \frac{d^2p}{(2\pi)^2}\frac{(p_u+k_{1u})(p_u-k_{2u})}{(p^2 + m^2-i\epsilon)((p+k_1)^2 + m^2-i\epsilon)((p-k_2)^2+m^2-i\epsilon)}\notag\, .
\end{align}
This is a triangle loop diagram.
Expanding at low momentum, 
\begin{align}\label{tttsep2d}
&\lll {\cal T}[ T_{uu}(k_3)\Theta(k_1)\Theta(k_2)]\rr_{\rm sep} = \frac{1}{6\pi}\left(k_{1u}^2 + 3k_{1u}k_{2u}+ k_{2u}^2\right)+O(k^4)\, .
\end{align}
We calculated the loop in \eqref{tttsep2dLoop} by combining the denominator with Feynman parameters, then expanding in $k$. Finally we evaluate the sum rule in the form \eqref{csumT}, 
\begin{align}
\Delta c &= 3\pi \int du_3 \int d^2 x_1 d^2 x_2\, u_1^2 \delta(u_2) \langle {\cal T}[ T_{uu}(u_3,v_3=0) \Theta(x_1) \Theta(x_2)] \rangle_{\rm sep} \\
&=  -3\pi  \left. (\p_{k_{1u}}-\p_{k_{2u}})^2 \lll  {\cal T}[T_{uu}(-k_1-k_2)\Theta(k_1)\Theta(k_2)] \rr_{\rm sep}  \right|_{k_1=k_2=0}
\notag\\
&= 1\, . \notag
\end{align}
This produces the correct $\Delta c$.

Note that the momentum dependence in \eqref{tttsep2d} does not naively match the CFT result \eqref{R32d}. It is only after setting $k_{3u} = 0$ that they agree. This reflects the presence of nonzero partial contact terms in the 3-point function, which drop out of the averaged null energy.

\subsubsection{Free massive Majorana fermion}

A free massive Majorana fermion in two dimensions flows from a massless Majorana fermion CFT (i.e. the 2d critical Ising model) with c$_{UV} = \frac{1}{2}$ to an empty theory in the IR with c$_{IR}=0$, such that $\Delta c=1/2$. The action was presented in \eqref{eq:FermionAction} while the $\braket{\psi(z,\bar{z})\psi(0)}$ propagator was presented in \eqref{eq:FermionProp1}. The two other fermion propagators are 
\begin{align}
 \braket{\bar{\psi}(z,\bar{z})\bar{\psi}(0,0)} &= 2\partial_{\bar{z}}\left[\int \frac{d^2p}{(2\pi)^2}\frac{e^{ip\cdot x}}{p^2 + m^2}\right] = \frac{1}{\pi}\partial_{\bar{z}}K_0\left(m|z|\right) =  -\frac{m}{2\pi}\left(\frac{z}{\bar{z}}\right)^{1/2}K_1\left(m|z|\right)\, ,\\
 \braket{\bar{\psi}(z,\bar{z})\psi(0,0)} &= -im \int \frac{d^2p}{(2\pi)^2}\frac{e^{p\cdot x}}{p^2 + m^2}  = -i\frac{m}{2\pi}K_0\left(m|z|\right)\label{eq:FermionProp2}\, .
\end{align}
Note that under the interchange of two fermions, all the correlators pick up a minus sign, i.e. 
\be 
\braket{\bar{\psi}(z,\bar{z})\bar{\psi}(0,0)}=-\braket{\bar{\psi}(0,0)\bar{\psi}(z,\bar{z})}\, ,\qquad \qquad 
\braket{\bar{\psi}(z,\bar{z})\psi(0,0)} = -\braket{\psi(\bar{z},z)\bar{\psi}(0,0)}\, .
\ee 
The stress tensor derived from the action \eqref{eq:FermionAction} is shown in \eqref{eq:TzzFermion}, and repeated here for convenience along with the trace: 
\be 
T_{zz}(z,\bar{z})=\frac{1}{2}\psi(z,\bar{z}) \partial_z\psi(z,\bar{z})\, ,\qquad \qquad \Theta = -im \bar{\psi}(z,\bar{z})\psi(z,\bar{z})\, .
\ee
The theory is free, so correlators can be computed using Wick's theorem.
The trace two point function is 
\be 
\braket{\Theta(z,\bar{z})\Theta(0,0)} = \left(\frac{m^2}{2\pi}\right)^2\left(K_1\left(m|z|\right)^2-K_0\left(m|z|\right)^2\right)\label{eq:TraceTraceFermion}
\ee
The Zamolodchikov sum rule is given in \eqref{zamosum}, and in radial coordinates, it becomes
\be 
\Delta c = 3\pi \int d^2x\, x^2\braket{\Theta(x)\Theta(0)}= 6\pi^2 \int_0^\infty dr\, r^3\braket{\Theta(r)\Theta(0)}\, ,\label{eq:ZamoRadial}
\ee 
with $r=\sqrt{z\bar{z}}$. Using \eqref{eq:TraceTraceFermion} within \eqref{eq:ZamoRadial}, we obtain 
\be 
\Delta c =  \frac{3}{2}m^4\int_0^\infty dr\, r^3\left(K_1\left(mr\right)^2-K_0\left(mr\right)^2\right)
=\frac{1}{2}\, .
\ee
Now that we have checked the $c$-theorem at the level of the trace-trace two-point function using Zamolodchikov's result, we can move to the sum rule involving the ANE operator. For this, we need to compute the three-point function. This is again a Wick contraction exercise, and we obtain
\begin{align}
&\braket{\mathcal{T}\left[T_{zz}(z_3,\bar{z}_3) \Theta(z_1,\bar{z}_1)\Theta(z_2,\bar{z}_2)\right]}_{\rm sep}=\\
&\qquad -\frac{m^2}{2}i\int \frac{d^2p_1}{(2\pi)^2}\frac{d^2p_2}{(2\pi)^2}\frac{d^2p_3}{(2\pi)^2}\frac{e^{ip_1\cdot x_{12} + i p_2\cdot x_{13} + i p_3 \cdot x_{23}}}{(p_1^2 + m^2-i\epsilon)(p_2^2 + m^2-i\epsilon)(p_3^2 + m^2-i\epsilon)}\nonumber\\
 &\phantom{=}\qquad\qquad  \times \left[2(p_{2z}-p_{3z})(m^2(p_{1z}-p_{2z}+p_{3z})+4p_{1\bar{z}}p_{2z}p_{3z})\right]\, , \notag
\end{align}
where $p_i\cdot x_{jk} = p_{iz}(z_j-z_k) + p_{i\bar{z}}(\bar{z}_j-\bar{z}_k)$. 
Going to momentum space, we obtain a one-loop integral that is performed by the same method as above. 
 We then use $k_{iz}\rightarrow -k_{iu}$ and $k_{i\bar{z}} \rightarrow k_{iv}$ and expand at low momenta to obtain
\be 
\lll \mathcal{T}\left[T_{uu}(k_3) \Theta(k_1)\Theta(k_2)\right]\rr_{\rm sep} =  \frac{k_{1u}^2+4 k_{1u}k_{2u}+k_{2u}^2}{24\pi}+ O(k^4)\, .
\ee 
We can now evaluate the sum rule in the form \eqref{csumT}. This amounts to 
\begin{align}
\Delta c &= 3\pi \int du_3 \int d^2 x_1 d^2 x_2\, u_1^2 \delta(u_2) \langle {\cal T}[ T_{uu}(u_3,v_3=0) \Theta(x_1) \Theta(x_2)] \rangle_{\rm sep} \\
&=  -3\pi  \left. (\p_{k_{1u}}-\p_{k_{2u}})^2 \lll  {\cal T}[T_{uu}(-k_1-k_2)\Theta(k_1)\Theta(k_2)] \rr_{\rm sep}  \right|_{k_1=k_2=0}
\notag\\
&= \frac{1}{2}\, , \notag
\end{align}
as expected.

\ \\

\noindent \textbf{Acknowledgments}
\noindent We thank Simon Caron-Huot, Horacio Casini, Jeevan Chandra, Clay Cordova, Diego Hofman, Austin Joyce, Denis Karateev, Murat Kologlu, Zohar Komargodski, Juan Maldacena, David Meltzer, Joao Penedones, Shu-Heng Shao, Nathan Seiberg, David Simmons-Duffin, and John Stout for helpful discussions. This work is funded by NSF grant PHY-2014071. GM is supported by the Simons Foundation grant 488649 (Simons
Collaboration on the Nonperturbative Bootstrap) and the Swiss National
Science Foundation through the project 200020\_197160 and through the
National Centre of Competence in Research SwissMAP. We also acknowledge support from NSF grant PHY-1748958 for participation in a KITP workshop. Part of this work was performed in part at Aspen Center for Physics, which is supported by NSF grant PHY-2210452.

\appendix

\section{Varying the trace anomaly \label{ap:TwoDim}}
In this appendix, we give more details on the CFT calculations necessary to obtain the results in the main text. By varying the conformal anomaly, we compute both Euclidean and Lorentzian two- and three-point functions involving the trace of the stress tensor. We will perform the variations in Euclidean spacetime, then Wick rotate to Lorentzian at the end.
These calculations (and our conventions) are similar to the four-dimensional CFT calculations in \cite{Hartman:2023qdn} and we refer the reader there for more detailed discussion of analytic continuations, choices made in the variational definitions, and the Ward identities.

\subsection{Correlator conventions\label{sec:CorrelatorConventions}}
Brackets $\braket{\cdots}$ denote connected correlators. In Euclidean signature, we define the stress tensor correlators through the variations
\begin{align}
\langle T_{\mu\nu}(x_1)\dots T_{\alpha\beta}(x_n)\rangle
&= \frac{(-2)^n}{\sqrt{g(x_1)}\dots\sqrt{g(x_n)}} \frac{\delta^n }{\delta g^{\mu\nu}(x_1)\dots \delta g^{\alpha\beta}(x_n)}\log Z\, .\label{eq:Tcorrelator1}\\
\langle \Theta(x_1)\dots \Theta(x_n)\rangle
&= \frac{(-2)^n}{\sqrt{g(x_1)}\dots\sqrt{g(x_n)}}
g^{\mu\nu}(x_1) \frac{\delta}{\delta g^{\mu\nu}(x_1)}
\cdots
g^{\alpha\beta}(x_n) \frac{\delta}{\delta g^{\alpha\beta}(x_n)} \log Z \,  \notag
\end{align}
and
\small
\begin{align}\label{mixedDefs}
\langle \Theta(x_1)T_{\mu\nu}(x_2)\rangle
&= \frac{4}{\sqrt{g(x_1)}\sqrt{g(x_2)}} \frac{\delta}{\delta g^{\mu\nu}(x_2)} g^{\alpha\beta}(x_1) \frac{\delta}{\delta g^{\alpha\beta}(x_1)} \log Z\\
\langle \Theta(x_1)\Theta(x_2)T_{\mu\nu}(x_3)\rangle
&= \frac{-8}{\sqrt{g(x_1)} \sqrt{g(x_2)} \sqrt{g(x_3)}} \frac{\delta}{\delta g^{\mu\nu}(x_3)}
g^{\alpha\beta}(x_2)\frac{\delta}{\delta g^{\alpha\beta}(x_2)} g^{\rho\sigma}(x_1) \frac{\delta}{\delta g^{\rho\sigma}(x_1)} \log Z \ . \notag
\end{align}
\normalsize
These conventions, in particular where to place the trace contractions and $\sqrt{g}$'s, affect the contact terms, and were chosen to simplify the discussion of partial contact terms in \cite{Hartman:2023qdn}.

\subsection{Trace anomaly}

The trace of the stress tensor in a 2d CFT is
\be 
\braket{\Theta(x)} = \frac{1}{24\pi}\left(c R(x)+ b\Lambda^2\right)\, ,\label{eq:2dtraceanom}
\ee
where $c$ is the central charge, $R$ is the Ricci scalar of the background, and the $b$ term is explained in section \ref{s:cfromAnec}. By varying the trace  \eqref{eq:2dtraceanom} according to the definitions in section \ref{sec:CorrelatorConventions}, we can obtain higher-point functions involving the stress tensor and its trace. 

\subsection{$\braket{T_{\mu\nu}\Theta}$}

To obtain the Euclidean two-point function, we vary the trace  \eqref{eq:2dtraceanom} once with respect to the background metric:
\be 
\braket{\Theta(x_1)T_{\mu\nu}(x_2)}  = -\frac{2}{\sqrt{g(x_1)}\sqrt{g(x_2)}}\frac{\delta}{\delta g^{\mu\nu}(x_2)}\left[\sqrt{g(x_1)}\braket{\Theta(x_1)}\right]\, .\label{eq:Var1ap1}
\ee
Using the variation of the Ricci scalar, we obtain 
\be 
\braket{\Theta(x_1)T_{\mu\nu}(x_2)}  = -\frac{1}{12\pi}\frac{1}{\sqrt{g(x_2)}}\left[-\frac{1}{2}g_{\alpha\beta}\left(cR + b\Lambda^2 \right)+c\left(R_{\alpha\beta}-\nabla_\alpha \nabla_\beta  + g_{\alpha\beta}\nabla^2\right)\right]\frac{\delta g^{\alpha\beta}(x_1)}{\delta g^{\mu\nu}(x_2)}\, .
\ee
We now use the variation
\be 
\frac{\delta g^{\alpha\beta}(x_1)}{\delta g^{\mu\nu}(x_2)} = \frac{1}{2}\left(\delta^\alpha_\mu \delta^\beta_\nu+ \delta^{\alpha}_\nu \delta^{\beta}_\mu\right)\tilde{\delta}^{(2)}(x_1-x_2)\, , \label{eq:VariationMetric}
\ee
The Dirac delta function in \eqref{eq:VariationMetric} is the density $\tilde{\delta}^{(d)}(x)$, which is related to the scalar Dirac delta function $\delta^{(d)}(x)$ by 
\be 
\delta^{(2)}(x) = \frac{\tilde{\delta}^{(2)}(x)}{\sqrt{g(x)}}\, ,
\ee
which obeys $\int d^2 x\sqrt{g}\delta^{(2)}(x) =1$. The Euclidean two-point function is thus
\be 
\braket{\Theta(x_1)T_{\mu\nu}(x_2)} =-\frac{1}{12\pi}\left[-\frac{1}{2}g_{\mu\nu}\left(cR + b\Lambda^2 \right)+c\left(R_{\mu\nu}-\nabla_{(\mu} \nabla_{\nu)}  + g_{\mu\nu}\nabla^2\right)\right]\delta^{(2)}(x_1-x_2)\, .\label{eq:TmunuThetaX}
\ee
In flat space, \eqref{eq:TmunuThetaX} becomes 
\be 
\braket{\Theta(x_1)T_{\mu\nu}(x_2)} =\frac{1}{12\pi}\left[\frac{1}{2}bg_{\mu\nu}\Lambda^2 +c\left(\partial_{\mu} \partial_{\nu}  - g_{\mu\nu}\partial^2\right)\right]\delta^{(2)}(x_1-x_2)\, ,
\ee
where all the derivatives act at position $x_1$, and $g_{\mu\nu}$ is the flat Euclidean metric. The Fourier transform to momentum space is
\be
\braket{\Theta(K_1)T_{\mu\nu}(K_2)} = \int d^2x_1 d^2x_2 \, e^{iK_1 x_1 + i K_2x_2}\braket{\Theta(x_1)T_{\mu\nu}(x_2)}\, ,
\ee
such that 
\be 
\lll \Theta(K_1)T_{\mu\nu}(-K_1)\rr = \frac{1}{12\pi}\left(\frac{1}{2}bg_{\mu\nu}\Lambda^2 -c\left(K_{1\mu} K_{1\nu}  - g_{\mu\nu}K_1^2\right)\right)\, ,\label{eq:TmunuTheta}
\ee
where the double bracket notation was defined in equation \eqref{eq:DBracket}. Note that the correlator in \eqref{eq:TmunuTheta} does depend explicitly on $b$ even in flat space.

It is straightforward to calculate the trace-trace two-point function from \eqref{eq:TmunuTheta}. 
We find 
\be 
\braket{\Theta(x_1)\Theta(x_2)} = g^{\mu\nu}\braket{\Theta(x_1)T_{\mu\nu}(x_2)}   = \frac{1}{12\pi}\left[-c\nabla^2 + b\Lambda^2 \right]\delta^{(2)}(x_1-x_2)\, ,
\ee
which is the curved spacetime expression. In flat space,
\be 
\braket{\Theta(x_1)\Theta(x_2)} = \frac{1}{12\pi}\left(-c\partial^2 + b\Lambda^2\right)\delta^{(2)}(x_1-x_2)\, .
\ee
This is \eqref{eq:EuclidenTraceTrace} in the main text. 
In momentum space, it becomes
\begin{align}
\lll\Theta(K_1)\Theta(-K_1)\rr
=& \frac{1}{12\pi}(c\K_1^2 + b\Lambda^2)\, ,\label{eq:B12}
\end{align}
which is the result \eqref{eq:TINv} of the main text.

\subsection{$\braket{T_{\mu\nu}\Theta\Theta}$}
With the conventions in \eqref{mixedDefs}, the three-point function can be calculated by the variation
\begin{align} 
\braket{\Theta(x_1)\Theta(x_2)T_{\mu\nu}(x_3)}& = -\frac{2}{\sqrt{g(x_1)}\sqrt{g(x_2)}\sqrt{g(x_3)}}\frac{\delta}{\delta g^{\mu\nu}(x_3)}\left[\sqrt{g(x_1)}\sqrt{g(x_2)}\braket{\Theta(x_1)\Theta(x_2)}\right]\notag\\
 &= \frac{4}{\sqrt{g(x_1)}\sqrt{g(x_2)}\sqrt{g(x_3)}}\frac{\delta}{\delta g^{\mu\nu}(x_3)}\left[g^{\alpha\beta}\frac{\delta}{\delta g^{\alpha\beta}(x_2)}\left[\sqrt{g(x_1)}\braket{\Theta(x_1)}\right]\right]\, .
\end{align}
We will use the following metric variation
\be 
\delta \left[\delta^{(2)}(x_{12}) \right]= \frac{1}{2}\delta^{(2)}(x_{12})g_{\alpha\beta}\delta g^{\alpha\beta}\, ,
\ee
where $x_{ij}=x_i-x_j$.
Performing these variations explicitly, and taking the flat space limit, we obtain the Euclidean three-point function 
\begin{align}
&\braket{\Theta(x_1)\Theta(x_2)T_{\mu\nu}(x_3)} =\frac{c}{6\pi}\left(\delta^{(2)}(x_{13})\partial_{\mu}\partial_{\nu}\delta^{(2)}(x_{12})+ \partial_{(\mu}\delta^{(2)}(x_{13})\partial_{\nu)}\delta^{(2)}(x_{12})\right)\label{eq:ToVary33AP}\\
&\qquad +\frac{g_{\mu\nu}}{12\pi}\left[ -c\left(\delta^{(2)}(x_{13})\partial^2\delta^{(2)}(x_{12})+\partial_\alpha \delta^{(2)}(x_{12}) \partial^\alpha \delta^{(2)}(x_{13})\right)+b \Lambda^2 \delta^{(2)}(x_{12})\delta^{(2)}(x_{13})\right]\, ,\notag
\end{align}
where all derivatives act at position $x_1$. The Fourier transform to momentum space is done by first converting the derivatives acting at $x_1$ to derivatives acting on $x_2$ and $x_3$, accounting for minus signs, which then makes Fourier transforming trivial. The result, for the Euclidean momentum space correlator is  
\begin{align}
&\lll\Theta(K_1)\Theta(K_2)T_{\mu\nu}(K_3)\rr =\frac{c}{6\pi}\left(K_{1(\mu}K_{2\nu)}\right) -\frac{g_{\mu\nu}}{12\pi}\left[c\left(K_1\cdot K_2 \right)-b \Lambda^2\right]\, .\label{eq:ToVary333AP}
\end{align}
In flat Euclidean space, the null-null component is 
\be 
\braket{\Theta(x_1)\Theta(x_2)T_{uu}(x_3)} = \frac{c}{6\pi}\left(\partial_{u_1} \delta^{(2)}(x_{12})\partial_{u_1} \delta^{(2)}(x_{13})+ \delta^{(2)}(x_{13})\partial_{u_1}^2\delta^{(2)}(x_{12})\right)\, ,
\ee
where we used $g_{uu}=0$.
All terms proportional to $b$ vanished, as needed for the sum rule. Finally, in momentum space, we obtain
\begin{align}
\lll\Theta(K_1)\Theta(K_2)T_{uu}(K_3)\rr 
=& \frac{c}{6\pi}K_{1u}K_{2u}\, ,\label{eq:3pt2d}\
\end{align}
which is the result \eqref{theta3e} in the main text. 

In \cite{Hartman:2023qdn}, we derived the Ward identities that follow from the definitions in section \ref{sec:CorrelatorConventions}. While we will not go through this exercice again here, we will just quote the equation that the three-point correlator must solve, which is 
\be 
K_3^\mu \lll T_{\mu\nu}(K_3)\Theta( K_1)\Theta(K_2)\rr  =- K_{2\nu}\lll \Theta(\
K_1)\Theta(-K_1)\rr - K_{1\nu}\lll \Theta(K_2)\Theta(-K_2)\rr\, .\label{eq:WIAp}
\ee 
Plugging in \eqref{eq:ToVary333AP} and \eqref{eq:B12}, it is easy to verify that this equation holds.

\subsection{Lorentzian correlators}
Following similar steps but in Lorentzian signature, or by Wick rotating the final answers, we obtain
\begin{align}
\lll {\cal T}[\Theta(k_1)\Theta(k_2)]\rr
&=-\frac{i}{12\pi }\left( c k_1^2 + b \Lambda^2\right)\, .
\end{align}
The Lorentzian three-point functions are 
\begin{align}
&\lll\mathcal{T}\left[\Theta(k_1)\Theta(k_2)T_{\mu\nu}(k_3)\right]\rr =-\frac{c}{6\pi}\left(k_{1(\mu}k_{2\nu)}\right) +\frac{g_{\mu\nu}}{12\pi}\left[c\left(k_1\cdot k_2 \right)-b \Lambda^2\right]\label{eq:ToVary333AP2}\\
&\lll\mathcal{T}\left[\Theta(k_1)\Theta(k_2)T_{uu}(k_3)\right]\rr =-\frac{c}{6\pi}k_{1u}k_{2u} \, .\label{eq:ToVary333AP3}
\end{align}

\addcontentsline{toc}{section}{References}
\bibliographystyle{utphys}
{\small
\bibliography{ref}

\providecommand{\href}[2]{#2}\begingroup\raggedright\begin{thebibliography}{10}

\bibitem{Zamolodchikov:1986gt}
A.~B. Zamolodchikov, ``{Irreversibility of the Flux of the Renormalization
  Group in a 2D Field Theory},'' {\em JETP Lett.} {\bfseries 43} (1986)
  730--732.

\bibitem{Cardy:1988cwa}
J.~L. Cardy, ``{Is There a c Theorem in Four-Dimensions?},''
  \href{http://dx.doi.org/10.1016/0370-2693(88)90054-8}{{\em Phys. Lett. B}
  {\bfseries 215} (1988) 749--752}.

\bibitem{Myers:2010tj}
R.~C. Myers and A.~Sinha, ``{Holographic c-theorems in arbitrary dimensions},''
  \href{http://dx.doi.org/10.1007/JHEP01(2011)125}{{\em JHEP} {\bfseries 01}
  (2011) 125},
\href{http://arxiv.org/abs/1011.5819}{{\ttfamily arXiv:1011.5819 [hep-th]}}.

\bibitem{Giombi:2014xxa}
S.~Giombi and I.~R. Klebanov, ``{Interpolating between $a$ and $F$},''
  \href{http://dx.doi.org/10.1007/JHEP03(2015)117}{{\em JHEP} {\bfseries 03}
  (2015) 117}, \href{http://arxiv.org/abs/1409.1937}{{\ttfamily arXiv:1409.1937
  [hep-th]}}.

\bibitem{Casini:2012ei}
H.~Casini and M.~Huerta, ``{On the RG running of the entanglement entropy of a
  circle},'' \href{http://dx.doi.org/10.1103/PhysRevD.85.125016}{{\em Phys.
  Rev. D} {\bfseries 85} (2012) 125016},
  \href{http://arxiv.org/abs/1202.5650}{{\ttfamily arXiv:1202.5650 [hep-th]}}.

\bibitem{Komargodski:2011vj}
Z.~Komargodski and A.~Schwimmer, ``{On Renormalization Group Flows in Four
  Dimensions},'' \href{http://dx.doi.org/10.1007/JHEP12(2011)099}{{\em JHEP}
  {\bfseries 12} (2011) 099}, \href{http://arxiv.org/abs/1107.3987}{{\ttfamily
  arXiv:1107.3987 [hep-th]}}.

\bibitem{Heckman:2015axa}
J.~J. Heckman and T.~Rudelius, ``{Evidence for C-theorems in 6D SCFTs},''
  \href{http://dx.doi.org/10.1007/JHEP09(2015)218}{{\em JHEP} {\bfseries 09}
  (2015) 218}, \href{http://arxiv.org/abs/1506.06753}{{\ttfamily
  arXiv:1506.06753 [hep-th]}}.

\bibitem{Cordova:2015vwa}
C.~Cordova, T.~T. Dumitrescu, and X.~Yin, ``{Higher derivative terms, toroidal
  compactification, and Weyl anomalies in six-dimensional (2, 0) theories},''
  \href{http://dx.doi.org/10.1007/JHEP10(2019)128}{{\em JHEP} {\bfseries 10}
  (2019) 128}, \href{http://arxiv.org/abs/1505.03850}{{\ttfamily
  arXiv:1505.03850 [hep-th]}}.

\bibitem{Cordova:2015fha}
C.~Cordova, T.~T. Dumitrescu, and K.~Intriligator, ``{Anomalies,
  renormalization group flows, and the a-theorem in six-dimensional (1, 0)
  theories},'' \href{http://dx.doi.org/10.1007/JHEP10(2016)080}{{\em JHEP}
  {\bfseries 10} (2016) 080}, \href{http://arxiv.org/abs/1506.03807}{{\ttfamily
  arXiv:1506.03807 [hep-th]}}.

\bibitem{Myers:2010xs}
R.~C. Myers and A.~Sinha, ``{Seeing a c-theorem with holography},''
  \href{http://dx.doi.org/10.1103/PhysRevD.82.046006}{{\em Phys.Rev.}
  {\bfseries D82} (2010) 046006},
\href{http://arxiv.org/abs/1006.1263}{{\ttfamily arXiv:1006.1263 [hep-th]}}.

\bibitem{cardyformula}
J.~L. Cardy, ``Operator content of two-dimensional conformally invariant
  theories,'' {\em Nucl. phys. B} {\bfseries 270} (1986) 186.

\bibitem{Holzhey:1994we}
C.~Holzhey, F.~Larsen, and F.~Wilczek, ``{Geometric and renormalized entropy in
  conformal field theory},''
  \href{http://dx.doi.org/10.1016/0550-3213(94)90402-2}{{\em Nucl.Phys.}
  {\bfseries B424} (1994) 443--467},
\href{http://arxiv.org/abs/hep-th/9403108}{{\ttfamily arXiv:hep-th/9403108
  [hep-th]}}.

\bibitem{Borde_1987}
A.~Borde, ``Geodesic focusing, energy conditions and singularities,''
  \href{http://dx.doi.org/10.1088/0264-9381/4/2/015}{{\em Classical and Quantum
  Gravity} {\bfseries 4} no.~2, (Mar, 1987) 343--356}.
  \url{https://doi.org/10.1088%2F0264-9381%2F4%2F2%2F015}.

\bibitem{Gao:2000ga}
S.~Gao and R.~M. Wald, ``{Theorems on gravitational time delay and related
  issues},'' \href{http://dx.doi.org/10.1088/0264-9381/17/24/305}{{\em Class.
  Quant. Grav.} {\bfseries 17} (2000) 4999--5008},
\href{http://arxiv.org/abs/gr-qc/0007021}{{\ttfamily arXiv:gr-qc/0007021
  [gr-qc]}}.

\bibitem{Graham_2007}
N.~Graham and K.~D. Olum, ``Achronal averaged null energy condition,''
  \href{http://dx.doi.org/10.1103/physrevd.76.064001}{{\em Physical Review D}
  {\bfseries 76} no.~6, (Sep, 2007) }.
  \url{https://doi.org/10.1103%2Fphysrevd.76.064001}.

\bibitem{Klinkhammer:1991ki}
G.~Klinkhammer, ``{Averaged energy conditions for free scalar fields in flat
  space-times},''
\href{http://dx.doi.org/10.1103/PhysRevD.43.2542}{{\em Phys. Rev.} {\bfseries
  D43} (1991) 2542--2548}.

\bibitem{Wald:1991xn}
R.~M. Wald and U.~Yurtsever, ``{General proof of the averaged null energy
  condition for a massless scalar field in two-dimensional curved
  space-time},''
\href{http://dx.doi.org/10.1103/PhysRevD.44.403}{{\em Phys. Rev.} {\bfseries
  D44} (1991) 403--416}.

\bibitem{Folacci:1992xg}
A.~Folacci, ``{Averaged null energy condition for electromagnetism in Minkowski
  space-time},''
\href{http://dx.doi.org/10.1103/PhysRevD.46.2726}{{\em Phys. Rev.} {\bfseries
  D46} (1992) 2726--2729}.

\bibitem{Ford:1995gb}
L.~H. Ford and T.~A. Roman, ``{Averaged energy conditions and evaporating black
  holes},'' \href{http://dx.doi.org/10.1103/PhysRevD.53.1988}{{\em Phys. Rev.}
  {\bfseries D53} (1996) 1988--2000},
\href{http://arxiv.org/abs/gr-qc/9506052}{{\ttfamily arXiv:gr-qc/9506052
  [gr-qc]}}.

\bibitem{Faulkner:2016mzt}
T.~Faulkner, R.~G. Leigh, O.~Parrikar, and H.~Wang, ``{Modular Hamiltonians for
  Deformed Half-Spaces and the Averaged Null Energy Condition},''
  \href{http://dx.doi.org/10.1007/JHEP09(2016)038}{{\em JHEP} {\bfseries 09}
  (2016) 038},
\href{http://arxiv.org/abs/1605.08072}{{\ttfamily arXiv:1605.08072 [hep-th]}}.

\bibitem{Hartman:2016lgu}
T.~Hartman, S.~Kundu, and A.~Tajdini, ``{Averaged Null Energy Condition from
  Causality},'' \href{http://dx.doi.org/10.1007/JHEP07(2017)066}{{\em JHEP}
  {\bfseries 07} (2017) 066},
\href{http://arxiv.org/abs/1610.05308}{{\ttfamily arXiv:1610.05308 [hep-th]}}.

\bibitem{Hofman:2008ar}
D.~M. Hofman and J.~Maldacena, ``{Conformal collider physics: Energy and charge
  correlations},'' \href{http://dx.doi.org/10.1088/1126-6708/2008/05/012}{{\em
  JHEP} {\bfseries 0805} (2008) 012},
\href{http://arxiv.org/abs/0803.1467}{{\ttfamily arXiv:0803.1467 [hep-th]}}.

\bibitem{Hofman:2009ug}
D.~M. Hofman, ``{Higher Derivative Gravity, Causality and Positivity of Energy
  in a UV complete QFT},''
  \href{http://dx.doi.org/10.1016/j.nuclphysb.2009.08.001}{{\em Nucl. Phys.}
  {\bfseries B823} (2009) 174--194},
\href{http://arxiv.org/abs/0907.1625}{{\ttfamily arXiv:0907.1625 [hep-th]}}.

\bibitem{Hartman:2015lfa}
T.~Hartman, S.~Jain, and S.~Kundu, ``{Causality Constraints in Conformal Field
  Theory},'' \href{http://dx.doi.org/10.1007/JHEP05(2016)099}{{\em JHEP}
  {\bfseries 05} (2016) 099},
\href{http://arxiv.org/abs/1509.00014}{{\ttfamily arXiv:1509.00014 [hep-th]}}.

\bibitem{Hartman:2016dxc}
T.~Hartman, S.~Jain, and S.~Kundu, ``{A New Spin on Causality Constraints},''
  \href{http://dx.doi.org/10.1007/JHEP10(2016)141}{{\em JHEP} {\bfseries 10}
  (2016) 141},
\href{http://arxiv.org/abs/1601.07904}{{\ttfamily arXiv:1601.07904 [hep-th]}}.

\bibitem{Hofman:2016awc}
D.~M. Hofman, D.~Li, D.~Meltzer, D.~Poland, and F.~Rejon-Barrera, ``{A Proof of
  the Conformal Collider Bounds},''
  \href{http://dx.doi.org/10.1007/JHEP06(2016)111}{{\em JHEP} {\bfseries 06}
  (2016) 111},
\href{http://arxiv.org/abs/1603.03771}{{\ttfamily arXiv:1603.03771 [hep-th]}}.

\bibitem{Cordova:2017zej}
C.~Cordova, J.~Maldacena, and G.~J. Turiaci, ``{Bounds on OPE Coefficients from
  Interference Effects in the Conformal Collider},''
  \href{http://dx.doi.org/10.1007/JHEP11(2017)032}{{\em JHEP} {\bfseries 11}
  (2017) 032},
\href{http://arxiv.org/abs/1710.03199}{{\ttfamily arXiv:1710.03199 [hep-th]}}.

\bibitem{Cordova:2018ygx}
C.~C\'ordova and S.-H. Shao, ``{Light-ray Operators and the BMS Algebra},''
  \href{http://dx.doi.org/10.1103/PhysRevD.98.125015}{{\em Phys. Rev. D}
  {\bfseries 98} no.~12, (2018) 125015},
\href{http://arxiv.org/abs/1810.05706}{{\ttfamily arXiv:1810.05706 [hep-th]}}.

\bibitem{Bautista:2019qxj}
T.~Bautista and H.~Godazgar, ``{Lorentzian CFT 3-point functions in momentum
  space},'' \href{http://dx.doi.org/10.1007/JHEP01(2020)142}{{\em JHEP}
  {\bfseries 01} (2020) 142}, \href{http://arxiv.org/abs/1908.04733}{{\ttfamily
  arXiv:1908.04733 [hep-th]}}.

\bibitem{Besken:2020snx}
M.~Be\c{s}ken, J.~De~Boer, and G.~Mathys, ``{On local and integrated
  stress-tensor commutators},''
  \href{http://dx.doi.org/10.1007/JHEP07(2021)148}{{\em JHEP} {\bfseries 21}
  (2020) 148}, \href{http://arxiv.org/abs/2012.15724}{{\ttfamily
  arXiv:2012.15724 [hep-th]}}.

\bibitem{Kelly:2014mra}
W.~R. Kelly and A.~C. Wall, ``{Holographic proof of the averaged null energy
  condition},'' \href{http://dx.doi.org/10.1103/PhysRevD.90.106003,
  10.1103/PhysRevD.91.069902}{{\em Phys. Rev.} {\bfseries D90} no.~10, (2014)
  106003}, \href{http://arxiv.org/abs/1408.3566}{{\ttfamily arXiv:1408.3566
  [gr-qc]}}.
[Erratum: Phys. Rev.D91,no.6,069902(2015)].

\bibitem{Afkhami-Jeddi:2016ntf}
N.~Afkhami-Jeddi, T.~Hartman, S.~Kundu, and A.~Tajdini, ``{Einstein gravity
  3-point functions from conformal field theory},''
  \href{http://dx.doi.org/10.1007/JHEP12(2017)049}{{\em JHEP} {\bfseries 12}
  (2017) 049},
\href{http://arxiv.org/abs/1610.09378}{{\ttfamily arXiv:1610.09378 [hep-th]}}.

\bibitem{Meltzer:2017rtf}
D.~Meltzer and E.~Perlmutter, ``{Beyond $a = c$: gravitational couplings to
  matter and the stress tensor OPE},''
  \href{http://dx.doi.org/10.1007/JHEP07(2018)157}{{\em JHEP} {\bfseries 07}
  (2018) 157},
\href{http://arxiv.org/abs/1712.04861}{{\ttfamily arXiv:1712.04861 [hep-th]}}.

\bibitem{Belin:2019mnx}
A.~Belin, D.~M. Hofman, and G.~Mathys, ``{Einstein gravity from ANEC
  correlators},'' \href{http://dx.doi.org/10.1007/JHEP08(2019)032}{{\em JHEP}
  {\bfseries 08} (2019) 032},
\href{http://arxiv.org/abs/1904.05892}{{\ttfamily arXiv:1904.05892 [hep-th]}}.

\bibitem{Kologlu:2019bco}
M.~Kologlu, P.~Kravchuk, D.~Simmons-Duffin, and A.~Zhiboedov, ``{Shocks,
  Superconvergence, and a Stringy Equivalence Principle},''
  \href{http://dx.doi.org/10.1007/JHEP11(2020)096}{{\em JHEP} {\bfseries 11}
  (2019) 096},
\href{http://arxiv.org/abs/1904.05905}{{\ttfamily arXiv:1904.05905 [hep-th]}}.

\bibitem{Baumann:2019ghk}
D.~Baumann, D.~Green, and T.~Hartman, ``{Dynamical Constraints on RG Flows and
  Cosmology},'' \href{http://dx.doi.org/10.1007/JHEP12(2019)134}{{\em JHEP}
  {\bfseries 12} (2019) 134}, \href{http://arxiv.org/abs/1906.10226}{{\ttfamily
  arXiv:1906.10226 [hep-th]}}.

\bibitem{Belin:2020lsr}
A.~Belin, D.~M. Hofman, G.~Mathys, and M.~T. Walters, ``{On the stress tensor
  light-ray operator algebra},''
  \href{http://dx.doi.org/10.1007/JHEP05(2021)033}{{\em JHEP} {\bfseries 05}
  (2021) 033}, \href{http://arxiv.org/abs/2011.13862}{{\ttfamily
  arXiv:2011.13862 [hep-th]}}.

\bibitem{Dixon:2019uzg}
L.~J. Dixon, I.~Moult, and H.~X. Zhu, ``{Collinear limit of the energy-energy
  correlator},'' \href{http://dx.doi.org/10.1103/PhysRevD.100.014009}{{\em
  Phys. Rev. D} {\bfseries 100} no.~1, (2019) 014009},
  \href{http://arxiv.org/abs/1905.01310}{{\ttfamily arXiv:1905.01310
  [hep-ph]}}.

\bibitem{Kologlu:2019mfz}
M.~Kologlu, P.~Kravchuk, D.~Simmons-Duffin, and A.~Zhiboedov, ``{The light-ray
  OPE and conformal colliders},''
  \href{http://dx.doi.org/10.1007/JHEP01(2021)128}{{\em JHEP} {\bfseries 01}
  (2021) 128},
\href{http://arxiv.org/abs/1905.01311}{{\ttfamily arXiv:1905.01311 [hep-th]}}.

\bibitem{Lee:2022ige}
K.~Lee, B.~Me\c{c}aj, and I.~Moult, ``{Conformal Colliders Meet the LHC},''
  \href{http://arxiv.org/abs/2205.03414}{{\ttfamily arXiv:2205.03414
  [hep-ph]}}.

\bibitem{Caron-Huot:2017vep}
S.~Caron-Huot, ``{Analyticity in Spin in Conformal Theories},''
  \href{http://dx.doi.org/10.1007/JHEP09(2017)078}{{\em JHEP} {\bfseries 09}
  (2017) 078},
\href{http://arxiv.org/abs/1703.00278}{{\ttfamily arXiv:1703.00278 [hep-th]}}.

\bibitem{Simmons-Duffin:2017nub}
D.~Simmons-Duffin, D.~Stanford, and E.~Witten, ``{A spacetime derivation of the
  Lorentzian OPE inversion formula},''
  \href{http://dx.doi.org/10.1007/JHEP07(2018)085}{{\em JHEP} {\bfseries 07}
  (2018) 085}, \href{http://arxiv.org/abs/1711.03816}{{\ttfamily
  arXiv:1711.03816 [hep-th]}}.

\bibitem{Hartman:2023qdn}
T.~Hartman and G.~Mathys, ``{Averaged Null Energy and the Renormalization
  Group},'' \href{http://arxiv.org/abs/2309.14409}{{\ttfamily arXiv:2309.14409
  [hep-th]}}.

\bibitem{PhysRevLett.60.2709}
J.~L. Cardy, ``Central charge and universal combinations of amplitudes in
  two-dimensional theories away from criticality,''
  \href{http://dx.doi.org/10.1103/PhysRevLett.60.2709}{{\em Phys. Rev. Lett.}
  {\bfseries 60} (Jun, 1988) 2709--2711}.
  \url{https://link.aps.org/doi/10.1103/PhysRevLett.60.2709}.

\bibitem{10.1007/BF02592679}
S.~Bernstein, ``{Sur les fonctions absolument monotones},''
  \href{http://dx.doi.org/10.1007/BF02592679}{{\em Acta Mathematica} {\bfseries
  52} no.~none, (1929) 1 -- 66}. \url{https://doi.org/10.1007/BF02592679}.

\bibitem{de07d37f-965f-344e-9946-a38128703ea8}
J.~D. Tamarkin, ``On a theorem of {S. Bernstein-Widder},'' {\em Transactions of
  the American Mathematical Society} {\bfseries 33} no.~4, (1931) 893--896.
  \url{http://www.jstor.org/stable/1989514}.

\bibitem{Cappelli:1990yc}
A.~Cappelli, D.~Friedan, and J.~I. Latorre, ``{C theorem and spectral
  representation},'' \href{http://dx.doi.org/10.1016/0550-3213(91)90102-4}{{\em
  Nucl. Phys. B} {\bfseries 352} (1991) 616--670}.

\bibitem{PhysRevD.15.463}
M.~Bander and C.~Itzykson, ``Quantum-field-theory calculation of the
  two-dimensional ising model correlation function,''
  \href{http://dx.doi.org/10.1103/PhysRevD.15.463}{{\em Phys. Rev. D}
  {\bfseries 15} (Jan, 1977) 463--469}.
  \url{https://link.aps.org/doi/10.1103/PhysRevD.15.463}.

\bibitem{Haag:1992hx}
R.~Haag, {\em {Local quantum physics: Fields, particles, algebras}}.
\newblock 1992.

\bibitem{Verch:1999nt}
R.~Verch, ``{The Averaged Null energy condition for general quantum field
  theories in two-dimensions},'' \href{http://dx.doi.org/10.1063/1.533130}{{\em
  J. Math. Phys.} {\bfseries 41} (2000) 206--217},
  \href{http://arxiv.org/abs/math-ph/9904036}{{\ttfamily
  arXiv:math-ph/9904036}}.

\bibitem{Komargodski:2011xv}
Z.~Komargodski, ``{The Constraints of Conformal Symmetry on RG Flows},''
  \href{http://dx.doi.org/10.1007/JHEP07(2012)069}{{\em JHEP} {\bfseries 07}
  (2012) 069}, \href{http://arxiv.org/abs/1112.4538}{{\ttfamily arXiv:1112.4538
  [hep-th]}}.

\bibitem{Meltzer:2021bmb}
D.~Meltzer, ``{Dispersion Formulas in QFTs, CFTs, and Holography},''
  \href{http://dx.doi.org/10.1007/JHEP05(2021)098}{{\em JHEP} {\bfseries 05}
  (2021) 098}, \href{http://arxiv.org/abs/2103.15839}{{\ttfamily
  arXiv:2103.15839 [hep-th]}}.

\bibitem{Epstein:1965zza}
H.~Epstein, V.~Glaser, and A.~Jaffe, ``{Nonpositivity of energy density in
  Quantized field theories},'' \href{http://dx.doi.org/10.1007/BF02749799}{{\em
  Nuovo Cim.} {\bfseries 36} (1965) 1016}.

\end{thebibliography}\endgroup
}

\end{document}